\documentclass[aps,nofootinbib,superscriptaddress, showpacs,preprintnumbers,  nofootinbibt,onecolumn]{revtex4-2}
\usepackage{epsfig}
\usepackage{multirow}
\usepackage{subcaption}
\usepackage{eurosym}
\usepackage{dcolumn}
\usepackage{rotating}
\usepackage{bm}
\usepackage{enumerate}
\usepackage{float}
\usepackage{epstopdf}
\usepackage{amsmath}
\usepackage{bm}
\usepackage{amsfonts}
\usepackage{amssymb}
\usepackage{graphicx}
\usepackage{alphalph,mathtools}
\usepackage{etoolbox}
\usepackage{color}
\usepackage[dvipsnames]{xcolor}
\usepackage{tensor}
\usepackage{amsthm}
\usepackage{booktabs}
\usepackage{hyperref}
\hypersetup{colorlinks,citecolor=blue}
\usepackage{footnote}
\usepackage{makecell,tabularx}

\def\be{\begin{equation}}
\def\ee{\end{equation}}
\def\bea{\begin{eqnarray}}
\def\eea{\end{eqnarray}}

\newcommand{\beq} {\begin{equation}}
\newcommand{\eeq} {\end{equation}}

\usepackage{relsize,exscale}
\usepackage{color, colortbl}

\definecolor{Gray}{gray}{0.9}
\definecolor{LightCyan}{rgb}{0.88,1,1}
\definecolor{lime}{HTML}{A6CE39}

\begin{document}

\title{\bf Cosmological tests of the dark energy models in Finsler-Randers Space-time}
\author{Z. Nekouee}
\email{zohrehnekouee@gmail.com}
\affiliation{School of Physics, Damghan University, Damghan, 3671641167, Iran}
\author{Himanshu Chaudhary}
\email{himanshuch1729@gmail.com}
\affiliation{Department of Applied Mathematics, Delhi Technological University, Delhi-110042, India}
\affiliation{Pacif Institute of Cosmology and Selfology (PICS), Sagara, Sambalpur 768224, Odisha, India}
\affiliation{Department of Mathematics, Shyamlal College, University of Delhi, Delhi-110032, India}
\author{S. K. Narasimhamurthy}
\email{nmurthysk@gmail.com}
\affiliation{Department of PG Studies and Research in Mathematics, Kuvempu University,\\ Jnana Sahyadri, Shankaraghatta-577 451, Shivamogga, Karnataka, India}
\author{S. K. J. Pacif}
\email{shibesh.math@gmail.com}
\affiliation{Pacif Institute of Cosmology and Selfology (PICS), Sagara, Sambalpur 768224, Odisha, India}
\author{Manjunath Malligawad}
\email{manjunathmalligawad91@gmail.com}
\affiliation{Department of PG Studies and Research in Mathematics, Kuvempu University,\\ Jnana Sahyadri, Shankaraghatta-577 451, Shivamogga, Karnataka, India}
\begin{abstract}
The Finsler-Randers space-time offers a novel perspective on cosmic dynamics, departing from the constraints of General Relativity. This paper thoroughly investigates two dark energy models resulting from the parametrization of $H$ within this geometric framework. We have conducted some geometrical and physical analysis of the dark energy models in Finslerian geometry. First, We have derived the field equations governing the universe's evolution within the Finsler-Randers formalism, incorporating the presence of dark energy. Through this, we explore its implications on cosmological phenomena, including cosmic expansion, late-time behavior of the universe, cosmological phase transition, and a few more. Also, we employ observational data such as Cosmic Chronometer, Supernovae, Gamma-Ray Bursts, Quasar, and baryon acoustic oscillations to constrain the parameters associated with dark energy in the Finsler-Randers universe. Comparing theoretical predictions with empirical observations, we assess the model viability and discern any deviations from the standard $\Lambda$CDM cosmology. Our findings offer intriguing insights into the nature of dark energy within this alternative gravitational framework, providing a deeper understanding of its role in shaping cosmic evolution. The implications of our results extend to fundamental cosmology, hinting at new avenues for research to unravel the mysteries surrounding dark energy and the geometric structure of the universe within non-standard gravitational theories.
\end{abstract}
\maketitle
\section{Introduction}\label{sec1}
Dark energy constitutes about 68 \% of the total energy content of the universe, a mysterious force thought to be responsible for the universe's accelerated expansion. There are numerous dark energy (DE)  models in the literature. However, the most straightforward and widely accepted candidate for DE is Einstein's cosmological constant $\Lambda$ \cite{Sahni2, Vishwakarma4, Peebles}.
The Hubble parameter (H) is an essential quantity in cosmology, representing the universe's expansion rate. The concept of cosmic acceleration was initially supported by the findings from observing distant supernovae of type Ia with high redshifts \cite{Perlmutter, Riess}. The motivation for the concept of acceleration stemmed from the independent observations conducted by the supernova teams directed by Perlmutter and Riess, employing diverse methodologies. These observations, along with data gathered from cosmic microwave background (CMB) \cite{Jaffe, Spergel} and studies on large-scale structures \cite{Bernardis, Hanany}, contributed to the development of the concept. Researchers provide strong evidence, indicating a more precise comprehension of the universe's accelerating expansion \cite{Riess2, Riess3, Weinberg, Astier, Amanullah}. The finding of cosmic acceleration has significant implications for physics and cosmology and has sparked investigations into the basic ideas that underlie cosmological theories. In recent years, many cosmological models have been proposed by researchers, either by changing the gravity theory itself or by modifying the energy-momentum tensor in Einstein's field equations (EFEs) \cite{Vishwakarma, Vishwakarma2, Vishwakarma3}. The development of EFEs played a pivotal role in the quest for precise solutions. As a direct outcome, the Schwarzschild exterior solution \cite{Schwarzschild} emerged as the inaugural exact solution. This solution was derived by incorporating the perfect fluid equation of state as a supplementary condition. Efforts in exact solutions development utilizing Einstein's field equations (EFEs) have been concentrated on specific cases, such as solutions for static and spherically symmetric metrics. Numerous noteworthy examples have been derived, including Einstein's static solution, Tolman's solutions \cite{Tolman}, the de-Sitter solution \cite{Sitter}, Adler's solutions \cite{Adlar}, Vaidya and Tikekar solution \cite{Vaidya}, Knutsen's solutions \cite{Knutsen}, Buchdahl's solution \cite{Buchdahl}, and Durgapal's solutions \cite{Durgapal}. These solutions represent a comprehensive exploration of the possibilities within the framework of EFEs, showcasing the versatility of the equations in capturing various physical scenarios. The observational data is a pivotal inclusion in the cosmology examination, enhancing the credibility and applicability of models. Another alternative avenue contributing substantially to modern cosmological investigations is using numerical computation.\\\\
In recent advancements in dark energy models, the integration of observational data has significantly improved accuracy in understanding the universe's expansion. A notable contribution to this field is S. K. J. Pacif's \cite{Pacif1} study, where two DE cosmological models are investigated. The study involved obtaining exact solutions of the EFEs through cosmological parametrization, specifically utilizing the Hubble parameter (H) within a flat FLRW background. J. K. Singh \cite{Singh} has investigated a cosmological model within the FLRW framework by employing the Energy Density Scalar Field Differential Equation (EDSFD) parametrization. Constraints derived from observational data are applied to a specific f(Q, T) gravity model \cite{Kale} characterized by the Hubble parameter transition. Numerical analysis is crucial in investigating the cubic parametrization of the deceleration parameter within the $f(R, T)$ gravitational theory \cite{Sofuoglu}. This study leverages observational data, specifically from $H(z)$ and SNIa datasets, to determine optimal model parameters and the Hubble constant through best-fit values. The brief analysis of three dark energy models, derived from deceleration parameter parametrization \cite{Dhruv}, reveals noteworthy traits like late-time acceleration and a cosmological transition from early deceleration to late acceleration. Abdulla A. M. \cite{Mamon} investigates the Barrow holographic dark energy model with the Granda–Oliveros infrared cutoff in the recent examination. Ashley Chraya, \cite{Chraya} uses the most recent available observational data to explore the cosmological constraints on the Variable Chaplygin gas model. Vinod Kumar Bhardwaj \cite{Vinod} conducted an investigation into an isotropic and homogeneous cosmological model of the universe within the framework of f(R, T) gravity. Recent studies have delved into various types of dark energy models, including those featuring nonmonotonic energy densities known as omnipotent dark energy (DE) models\cite{Shahnawaz}. Additionally, investigations have been conducted on the Horava-Lifshitz gravity model \cite{Madhur}.\\\\
In modern theoretical physics, the main goal is to explain how the universe behaves and understand its mysterious components differently. In the last 30 years, observations have indicated that the universe is extending at an accelerated rate, and standard FRW cosmology with ordinary matter fields (cold dark matter, radiation) is not to describe this.  The introduction of a cosmological constant $\Lambda$ in the classical Hilbert action of general relativity (GR) can explain this accelerated expansion. However, its value cannot be calculated theoretically and must be adjusted by hand to match observations.
Attempts have been made to relate the cosmological constant to the vacuum energy of quantum fields, but the calculated value deviates from the value suggested by observations.
The causality for the accelerated expansion is called "dark energy," although we know very little about it. Several modified gravity theories have been offered as nominees to explain theoretically the observations, usually by modifying the Hilbert action in a Riemannian framework. Finsler Geometry is a hopeful approach to unraveling the secrets of gravity more comprehensively. Instead of using pseudo-Riemannian geometry to describe the gravitational interaction, Finsler space-time geometry offers a viable method for a geometric explanation of the dark matter and dark energy phenomena \cite{Bogoslovsky, Gibbons, Kostelecky, Pfeifer, Lammerzahl}.
Within the context of Finsler geometry, various mathematical expressions can be studied, e.g., cosmological implications of scalar-tensor theories were studied in Ref. \cite{Ikeda}.
These theories were obtained, in effect, from the Lorentz fiber bundle of a Finsler-like geometry.
In recent studies, some researchers have investigated dark energy models and used the Finsler geometry as their framework of choice.
Finsler and Finsler-like theories take a different approach because they change the geometry instead of the action.
Generally, metrical extensions of Riemann geometry can be equipped with a Finsler geometrical structure in a manifold that leads to generalized gravitational field theories. They are drawing inspiration from the osculating Barthel–Kropina geometry \cite{Bouali, Hama}, Barthel–Randers geometry \cite{Hama2}, Finsler–Randers geometry \cite{Wenyu, Han} in their investigation. During these years, applications of Finsler geometry, especially the Finsler-Randers (F-R) model, have developed rapidly, mainly in general relativity, astrophysics, and cosmology. The F-R field equations supply a Hubble parameter that contains an extra geometrical term that can be utilized as a probable nominee for dark energy \cite{Stavrinos}.\\\\
On the other hand, recent observations of the cosmic microwave background (CMB) by experimentations like the Planck satellite have demonstrated potential anomalies and deviations from statistical isotropy on large angular scales \cite{441,442}. Noteworthy studies have been executed on the observable anisotropies of the universe \cite{45,46,47,48}.
These are associated with the very early state of the universe and related to the estimates of the Wiener-filtered map of the cosmic microwave background (CMB), the anisotropic pressure, or the composition of a primordial vector field (instance magnetic field) to the metrical spatial structure of the universe \cite{49,50,51}.
In this case, the form of the scale factor can be impacted by the preliminary field. Lately, anisotropies also have been discussed by the Royal Society, and there is a compelling case that there are anomalies in data \cite{511}. These anomalies, including the well-known low multipole alignments, cold spot, hemispherical asymmetry, and parity asymmetry, have sparked expansive discussions within the cosmology community about their origins and implications. While these anomalies could be mere statistical flukes within the standard $\Lambda$-CDM model, they have also been interpreted as potential signatures of new physics or extensions to the standard cosmological framework.
One fascinating possibility is that these anomalies might be related to the primordial anisotropies' existence or preferred directions in the early universe. Such anisotropies could derive from various sources, such as anisotropic inflationary models, vector or tensor fields, or even more exotic scenarios like cosmic birefringence or violations of fundamental symmetries. Probing the effects of such anisotropies on the evolution of the universe and the formation of large-scale structures is an active area of research with potential importance for our understanding of the early universe and the nature of dark energy.\\\\
A geometry that possibly connects the Riemannian metric structure of space-time to physical vector fields is a class of the Finsler-Randers variousness of spaces. In these spaces, an electromagnetic field, a magnetic field, or a gauge vector field may derive from the universe's physical source
and can be contained in the geometry and create an anisotropic structure \cite{52,53,54,55,56}. As a generalization of Riemannian geometry, Finsler geometry \cite{Bao,58} is the most general geometry where the line element depends on space-time coordinates (x) and tangent vectors (y).
This y-dependence essentially describes the Finsler field and is merged with the concept of anisotropy, which creates deviations from Riemannian geometry \cite{59}.
Therefore, internal variables generate these geometrical anisotropies. With these conditions, Finsler geometry can be regarded as a physical geometry on which the matter dynamics occur, while Riemann geometry is the gravitational geometry \cite{60,61, Vacaru}.\\
\section{Gravitational Field Equations for Finsler-Randers Metric}
The F-R cosmic scenario is established on the Finalerian geometry, which expands the Riemannian geometry. It is worth noticing that Riemannian geometry is also the particular case of the Finslerian geometry. Here, we examine only the main characteristics of the theory (see \cite{Bao,Vacaru,Rund}). Generally, a Finsler space is derived from a generating function F(x, y) on the tangent bundle TM of a manifold M. The generating function F is differentiable on $\widehat{TM} = TM\setminus\{0\}$. The function F is also positively homogeneous of degree one in $y$. The $n\times n$ Hessian matrix is positive definite at every point of $\widehat{TM}$ that as the following form,
\begin{equation}\label{Eq1}
		g_{ij}=\frac{\partial^{2}(\frac{1}{2}F^2)}{\partial y^{\mu}\partial y^{\nu}}=\frac{1}{2}\dot{\partial}_{\mu}\dot{\partial}_{\nu}F^{2},
	\end{equation}
In other words, pair ($M, F$) is called Finsler space-time. In the F-R space-time, we have
\begin{equation}\label{Eq2}
F=\alpha(x,y)+\beta(x,y),
\end{equation}
where
\begin{equation}\label{Eq3}
\alpha(x,y)=\sqrt{a_{ij}y^iy^j}, ~~~~~ \beta(x,y)=\mathfrak{u}_{i}y^{i},
\end{equation}
$a_{ij} (x)$ is the FRW metric defined as a tensor corresponding to the Riemannian metric,
\begin{equation}\label{Eq4}
a_{ij}(x)=diag\left(1,-\frac{a^2(t)}{1-\kappa r^2},-a^2(t)r^2,-a^2(t)r^2\sin^2\theta\right),
\end{equation}
and $\kappa = 0,\pm1$ for a flat, closed and hyperbolic geometry respectively.
Consequently, the geometrical structure of Finsler-Randers space is a generalization of Riemannian geometry.
Here, the spatial coordinates are comoving the time coordinate defines the proper time investigated by the comoving observer. The vector $y^i = \frac{dx^i}{d\tau}$ expresses the four tangent vectors that depict the velocity of a comoving observer towards an appropriate family of world lines, namely fluid flow pathways, in a locally anisotropic universe. The arclength $\tau$ is the proper time. So we have $y^i=(1, 0, 0, 0)$.\\
$\mathfrak{u}_i$ expresses a weak primordial vector field with $|\mathfrak{u}_i|\ll1$ and $\mathfrak{u}_i = (\mathfrak{u}_0, 0, 0, 0)$, where $\mathfrak{u}_0=\mathfrak{u}_0(t )$. Here, $\beta$ is an 1-form.
Now, using Eq. (\ref{Eq1}), the metric for the Finslerian space-time can be defined as follows \cite{Stavrinos,Triantafyllopoulos}:
\begin{equation}\label{Eq5}
g_{ij}=\mathfrak{g}_{ij}+\frac{1}{4\alpha}\left(\mathfrak{u}_iy_j+\mathfrak{u}_jy_i\right)-\frac{\beta}{\alpha^3}y_iy_j+\mathfrak{u}_i\mathfrak{u}_j,
\end{equation}
where
\begin{equation}\label{Eq6}
\mathfrak{g}_{ij}=\frac{F}{\alpha}(x,y)a_{ij}(x).
\end{equation}
Cartan tensor can be easily defined in full expression as follows \cite{Raushan1}
\begin{equation}\label{Eq7}
C_{ijk}=\frac{1}{2}\left[\frac{1}{\alpha}\mathfrak{S}_{(ijk)}(a_{ij}\mathfrak{u}_{k})-\frac{1}{\alpha^3}\mathfrak{S}_{(ijk)}(y_iy_j\mathfrak{u}_{k})-
\frac{\beta}{\alpha^3}\mathfrak{S}_{(ijk)}(a_{ij}y_{k})+\frac{3\beta}{\alpha^5}y_iy_jy_k\right],
\end{equation}
where $\mathfrak{S}_{(ijk)}$ expresses the sum over the cyclic permutation of the indices. It has been found that $C_{000}=\frac{\mathfrak{u}_{0}}{2}$ \cite{Stavrinos}.\\
The Finsler–Randers field equations are the following form,
\begin{equation}\label{Eq8}
L_{ij}=\frac{8\pi G}{c^4}\left(T_{ij}-\frac{1}{2}T\mathfrak{g}_{ij}\right),
\end{equation}
where
\begin{itemize}
  \item $T_{ij}$ are the energy-momentum tensor components for describing Finslerian anisotropic matter
\begin{equation}\label{Eq9}
T_{ij}(x,y(x))=(\rho+P)y_{i}(x)y_{j}(x)-P\mathfrak{g}_{ij}(x,y(x)),
\end{equation}
and $T$ is the trace of them. Thus we have
\begin{equation}\label{Eq10}
T_{00}=\rho, ~~~ T_{11}=-P \mathfrak{g}_{11}, ~~~ T_{22}=-P \mathfrak{g}_{22}, ~~~ T_{33}=-P \mathfrak{g}_{33}, ~~~ T=\rho\frac{\alpha}{F}-3P,
\end{equation}
where $\frac{F}{\alpha}\approx 1$ at the weak field limit.
\item $L_{ij}=L^{k}_{ikj}$ is Ricci tensor and by inflicting the conditions $\ddot{\mathfrak{u}}_0,\dot{\mathfrak{u}}_{0}^{2}\approx0$, its non-zero components as the following form \cite{Stavrinos2},
\begin{align}
  L_{00}= & -3\left(\frac{\ddot{a}}{a}+\frac{\dot{a}}{a}\dot{\mathfrak{u}}_{0}\right), \label{Eq11}\\
  L_{11}= & \frac{a\ddot{a}+2\dot{a}^2+2\kappa+4a\dot{a}\dot{\mathfrak{u}}_{0}}{1-\kappa r^2}, \label{Eq12}\\
  L_{22}= & \left(a\ddot{a}+2\dot{a}^2+2\kappa+4a\dot{a}\dot{\mathfrak{u}}_{0}\right)r^2, \label{Eq13}\\
  L_{33}= & \left(a\ddot{a}+2\dot{a}^2+2\kappa+4a\dot{a}\dot{\mathfrak{u}}_{0}\right)r^2\sin^2\theta. \label{Eq14}
\end{align}
\end{itemize}
Now by imposing $c=1$ and $\kappa=0$ we have
\begin{eqnarray}
\frac{\ddot{a}}{a}+2\frac{\dot{a}}{a}\dot{\mathfrak{u}}_{0}=\frac{-4\pi G}{3}(\rho+3P),\hspace{0.5cm}\label{Eq15}\\
\frac{\ddot{a}}{a}+2\frac{\dot{a}^{2}}{a^{2}}+4\frac{\dot{a}}{a}\dot{\mathfrak{u}}_{0}=4\pi G(\rho-P),\label{Eq16}
\end{eqnarray}
where the symbol dot denotes the derivative with respect to t and $\dot{\mathfrak{u}}_{0}<0$ \cite{Roopa}.
By using Eqs. (\ref{Eq15}) and (\ref{Eq16}) we obtain the Friedmann equation
\begin{equation}\label{Eq17}
\frac{\dot{a}^{2}}{a^{2}}+\frac{\dot{a}}{a}\dot{\mathfrak{u}}_{0}=\frac{8\pi G}{3}\rho,
\end{equation}
and
\begin{equation}\label{Eq18}
-2\frac{\ddot{a}}{a}-\frac{\dot{a}^{2}}{a^{2}}-5\frac{\dot{a}}{a}\dot{\mathfrak{u}}_{0}=8\pi G P.
\end{equation}
Eq. (\ref{Eq17}) is similar to the Friedmann equation in the Riemannian framework, apart from the extra term $\frac{\dot{a}}{a}\dot{\mathfrak{u}}_{0}$. We associate this extra term with the present universe’s anisotropy. By using Eqs. (\ref{Eq17}) and (\ref{Eq18}) we obtain deformed parameters as following form
\begin{eqnarray*}
\tilde{H}^2&=&\frac{\dot{a}^{2}}{a^{2}}+\frac{\dot{a}}{a}\dot{\mathfrak{u}}_{0}, ~ \xrightarrow{\dot{\mathfrak{u}}_{0}^{2}\approx 0} ~ \tilde{H}=\left(\frac{\dot{a}}{a}+\frac{\dot{\mathfrak{u}}_{0}}{2}\right),\\ \tilde{q}&=&-\left(\frac{\ddot{a}}{a}+\frac{\dot{a}}{a}\dot{\mathfrak{u}}_{0}\right)\tilde{H}^{-2},
\end{eqnarray*}
where $\tilde{H}$ and $\tilde{q}$ are Hubble parameter and deceleration parameter, respectively.
Conservation of energy-momentum results from Eqs. (\ref{Eq17}) and (\ref{Eq18}) as follows,
\begin{equation}\label{Eq20}
8\pi G \dot{\rho}+24\pi G (\rho+P)H+6H^2\dot{\mathfrak{u}}_{0}+12\pi G(\rho+P)\dot{\mathfrak{u}}_{0}=0.
\end{equation}
The continuity Eq (\ref{Eq20}) plays an essential role in evolution because it deals with matter and its interactions. In current cosmology, two types of dark energy models are generally debated.
Dark energy interaction models (taking into account the interaction between cold dark matter and dark energy) \cite{Zimdahl} and non-interaction models of dark energy, where matters are authorized to evolve individually \cite{Raushan}. Up to date, there are no known interactions other than gravity between the matter and dark energy. The present study only refers to non-interacting models. The system of equations is non-linear ordinary differential equations, and finding exact solutions is difficult.
In the past, many attempts have been made to find exact and numerical solutions for EFEs. In the next section, solving techniques from the above system of equations will be discussed in detail.
\section{Einstein Field Equations for Finsler-Randers Metric with Dark Energy}
The nature of dark energy and its candidature are not yet known, and its expression as a source term in Einstein's field equations is speculative. However, DE is assumed to be homogeneous, permeating all over the space, and the energy-momentum tensor can be expressed as a perfect fluid  $T_{ij}^{\mathfrak{de}}$. In this case, its equation of state is in the form $P_{\mathfrak{de}}=\omega_{\mathfrak{de}}\rho_{\mathfrak{de}}$ where $\omega_{\mathfrak{de}}$ is the equation of state (EoS) parameter and is a function of time.
Generally, it holds for the inequality $\omega_{\mathfrak{de}} < 0$.
The appropriate value of $\omega_{\mathfrak{de}}$ is hotly debated, and analysis of some observational data suggests that its value is $-1.018\pm 0.031$ (Planck 2018 or SNe $IA^{a}$) \cite{Oikonomou}. Different values of $\omega_{\mathfrak{de}}$ in particular ranges produce different candidates, and they can be generally classified as seen in Table \ref{tbl1},
\begin{table*}
\begin{center}
\begin{tabular}{|l |c|c|}
\hline
Parameter    & Classification    \\
\hline
   $\omega_{\mathfrak{de}}=-1$ & \text{cosmological constant}\\
\hline
 $\omega_{\mathfrak{de}}=$ \text{Constant}$ \neq-1$ & \text{cosmic strings, domain walls, etc.}\\
\hline
   $\omega_{\mathfrak{de}} \neq $\text{Constant} & \makecell{scalar fields (quintessence, k-essence etc.), \\ braneworlds, Dirac-Born-Infeld (DBI)
action, Chaplygin gas etc.}\\
\hline
 $\omega_{\mathfrak{de}}<-1$ & \text{phantom models}\\
\hline
\end{tabular}
\caption{Classification of $\omega_{de}$ parameter.}\label{tbl1}
\end{center}
\end{table*}
See \cite{Copeland,Bamba} (and Refs. therein) for a more extensive list of dark energy models.
In each of them, fascinating cases exist that are associated with problems. For example, the cosmological constant is the most consistent model for dark energy that explains the observations but faces fine-tuning issues.
Similarly, phantom models are fascinating, where the weak energy condition ($\rho > 0$,  $P + \rho > 0$) with finite time singularity property is violated. In general relativity, dark energy can be introduced by adding the energy-momentum tensor $T^{\mathfrak{de}}$ as a perfect fluid together with the matter source $T^{\mathfrak{m}}$ into the Einstein field equations $G_{ij} = 8\pi GT_{ij}$ as following form,
\begin{equation}\label{Eq21}
T^{\mathfrak{tot}}=T^{\mathfrak{de}}+T^{\mathfrak{m}},
\end{equation}
$\rho_{\mathfrak{tot}}=\sum\rho+\rho_{\mathfrak{de}} $ and $P_{\mathfrak{tot}}=\sum P+P_{\mathfrak{de}}$ denote total energy densities and the total pressure corresponding to all kinds of matter (baryonic matter, dark matter, and radiation) and dark energy, respectively.
According to Eq. (\ref{Eq21}), we can rewrite Eqs. (\ref{Eq17}), (\ref{Eq18}), and (\ref{Eq20}) as follows,
\begin{eqnarray}
\frac{\dot{a}^{2}}{a^{2}}+\frac{\dot{a}}{a}\dot{\mathfrak{u}}_{0}=\frac{8\pi G}{3}\rho_{\mathfrak{tot}},\hspace{3cm}\label{Eq22}\\
-2\frac{\ddot{a}}{a}-\frac{\dot{a}^{2}}{a^{2}}-5\frac{\dot{a}}{a}\dot{\mathfrak{u}}_{0}=8\pi G P_{\mathfrak{tot}},\hspace{2.4cm}\label{Eq23}\\
8\pi G \dot{\rho}_{\mathfrak{tot}}+24\pi G (\rho_{\mathfrak{tot}}+P_{\mathfrak{tot}})H+6H^2\dot{\mathfrak{u}}_{0}+12\pi G(\rho_{\mathfrak{tot}}+P_{\mathfrak{tot}})\dot{\mathfrak{u}}_{0}=0.\label{Eq24}
\end{eqnarray}
The system of the above equations has only two independent equations with five unknowns $a$, $\rho$, $P$, $\rho_{\mathfrak{de}}$, $P_{\mathfrak{de}}$ (or $\omega_{\mathfrak{de}}$). The universe's large-scale homogeneous distribution of matter requires the barotropic equation of state to be considered $P=\omega\rho$, $\omega\in [0, 1]$. The equation of state expresses various kinds of matter sources in the universe depending on the discrete or dynamical values of the equation of state parameter as follows: baryonic matter ($\omega_{\mathfrak{bm}}= 0$), dark matter ($\omega_{\mathfrak{dm}}= 0$), radiation ($\omega_{\mathfrak{r}}= \frac{1}{3}$), stiff matter ($\omega_{\mathfrak{s}}= 1$), quintessence ($-1<\omega_{\mathfrak{q}}< -\frac{1}{3}$), etc. \\\\
This extra equation represents the third restriction equation. Another restriction equation can be the regard of the EoS of dark energy ($\omega_{\mathfrak{de}}=$ constant or a function of time $t$ or function of the scale factor or redshift), which is well known as a parametrization of dark energy EoS. These four equations can describe the cosmological dynamics of the universe where all the geometrical or physical parameters such as Hubble parameter ($H$), deceleration parameter ($q$), jerk parameter ($j$), densities ($\rho$ , $\rho_{\mathfrak{de}}$), pressures ($P$, $P_{\mathfrak{de}}$), EoS parameter ($\omega_{\mathfrak{de}}$), density parameter ($\Omega_{i}$), etc. are signified as functions of either scale factor or the redshift $z=\frac{a_{0}}{a}-1$  ($a_{0}$ is the current value of the scale factor, usually normalized to $a_{0} = 1$). But, we still need another equation to solve the system. The time evolution of the scale factor a is not yet known.
Various schemes exist for parametrizing the scale factor and its higher-order derivatives (H, q, j, etc.).
They provide the complete solution of EFEs, i.e., the explicit forms of the cosmic parameters as a function of $t$.
An essential analysis of the solution methods of EFEs in general relativity theory or modified theories has two parts:
One is the parametrization of geometrical parameters, and another is the parametrization of the physical parameters.\\
The first type of parameterization scheme (geometric parameterization) is considered to find exact solutions that discuss the expansion dynamics of the universe and provide the time evolution of the physical parameters. This method is generally known as an independent model for cosmological models or cosmological parameterization \cite{Cunha,Mortsell,Pacif}. It does not affect the background theory and provides solutions for EFEs.
As well as it has the advantage of reconstructing the cosmological history of the universe and describing some of the phenomena of the universe.
The goal is to achieve the exact solution of the Einstein field equations in standard general relativity with a simple parametrization of the Hubble parameter under the Finsler-Randers structure.
\section{Parametrization of Hubble Parameter in Finsler-Randers Model}
Cosmological investigations provide clues to explore the evolution of the observable universe in a model-independent way in terms of kinematic variables \cite{Capozziello}. Furthermore, the cosmological parameters analysis helps to study dark energy without any assumptions of any particular cosmological model other than the cosmological principle.
In this paper, we intend to use a simple parametrization of the Hubble parameter, which is an explicit function of the cosmic time “t” as following form \cite{Pacif},
\begin{equation}\label{Eq25}
H(t)=\frac{k_2t^{\mathfrak{m}}}{\left(t^{\mathfrak{n}}+k_1\right)^{\mathfrak{p}}},
\end{equation}
where $\mathfrak{m}, \mathfrak{n}, \mathfrak{p}\in \mathbb{R}$, $k_1$, $k_2\neq0$, and both have the dimensions of time. Some special values of the parameters $\mathfrak{m}$, $\mathfrak{n}$, and $\mathfrak{p}$ present some distinct models \cite{Pacif}. The parameterization (\ref{Eq25}) generalizes several well-known models, e.g., $\Lambda$CDM model, power law model, hybrid expansion model, jump model, linear variable deceleration parameter model, etc.
As we know, $|\mathfrak{u}_i|\ll1$ for parametrization Hubble parameter in Finsler structure we take $\mathfrak{u}_{0}=\exp(-t)$.
Table \ref{tbl2}  can describe cosmic phase transition phenomena for negative values of $k_1$, and $k_2$ in two models.
\begin{table*}[hpt]
\begin{center}
\begin{tabular}{|l |c|c|c|c|}
\hline
Models    & $H(t)$ & $\tilde{H}(t)$ & $a(t)$ & $\tilde{a}(t)$   \\
\hline
   \text{Model 1} & $\frac{k_{2}}{t(k_1-t)}$ & $\frac{k_{2}}{t(k_1-t)}-\frac{1}{2}e^{-t}$ & $\beta\left(\frac{t}{k_1-t}\right)^{\frac{k_2}{k_1}}$ & $\beta\left(\frac{t}{k_1-t}\right)^{\frac{k_2}{k_1}}e^{\frac{1}{2}e^{-t}}$\\
\hline
\text{Model 2} & $\frac{k_{2}}{t(k_1-t^2)}$ & $\frac{k_{2}}{t(k_1-t^2)}-\frac{1}{2}e^{-t}$ & $\beta\left(\frac{t^2}{k_1-t^2}\right)^{\frac{k_2}{2k_1}}$ & $\beta\left(\frac{t^2}{k_1-t^2}\right)^{\frac{k_2}{2k_1}}e^{\frac{1}{2}e^{-t}}$\\
\hline
\end{tabular}
\caption{The Models}\label{tbl2}
\end{center}
\end{table*}
In studying late-time universe and observational studies, expressing all the cosmological parameters as functions of redshift z is convenient. It is well known that the Hubble parameter ($H =\frac{\dot{a}}{a}$) directly probes the universe's expansion history where $\dot{a}$ is the rate of change of the universe's scale factor.
The Hubble parameter is also related to the differential redshift $H(z) =\frac{1}{z+1}\frac{dz}{dt}$, where $dz$ is obtained from spectroscopic surveys. Table \ref{tblH} shows the t-z relations, that $\alpha=\frac{k_1}{k_2}$ and parameters $\alpha$ and $\beta$ are sufficient to describe these models.\\
\begin{table*}[hpt]
\centering
\begin{tabular}{|l |c|c|}
\hline
Models    & $t(z)$ & $\tilde{H}(z)$     \\
\hline
   \text{Model 1} & $\frac{k_1}{1+\left[\beta(1+z)\right]^\alpha}$ & $\frac{H_0\left(1+\left[\beta(1+z)\right]^\alpha\right)^2}{\left(1+\beta^\alpha\right)^2\left(1+z\right)^\alpha}
   -\frac{1}{2}\rm{e}^{-\frac{\left(1+\beta^\alpha\right)^2}{\alpha H_0\beta^\alpha\left(1+\left[\beta(1+z)\right]^\alpha\right)}}$\\
\hline
\text{Model 2} & $\frac{\sqrt{k_1}}{\sqrt{1+\left[\beta(1+z)\right]^{2\alpha}}}$ &$\frac{H_0\left(1+\left[\beta(1+z)\right]^{2\alpha}\right)^{\frac{3}{2}}}{\left(1+\beta^{2\alpha}\right)^\frac{3}{2}\left(1+z\right)^{2\alpha}}
-\frac{1}{2}\rm{e}^{-\frac{\left(1+\beta^{2\alpha}\right)^\frac{3}{2}}{\alpha H_0\beta^{2\alpha}\left(1+\left[\beta(1+z)\right]^{2\alpha}\right)^\frac{1}{2}}}$ \\
\hline
\end{tabular}
\caption{Hubble parameter vs. z in two models.}\label{tblH}
\end{table*}

\section{Methodology}
In cosmology, parameter estimation often employs the Bayesian framework, which entails calculating the posterior distribution of parameters \(\theta\) given the observed data \(D\). This relationship is represented as:
\begin{equation}
P(\theta \mid D) = \frac{\mathcal{L}(D \mid \theta) P(\theta)}{P(D)},
\end{equation}
where \(\mathcal{L}(D \mid \theta)\) is the likelihood function, \(P(\theta)\) denotes the prior distribution of the parameters, and \(P(D)\) is the marginal likelihood. The posterior distribution is explored across the parameter space \(\theta\), typically using algorithms such as Metropolis-Hastings. This algorithm directs a random walker through the parameter space, favoring areas with higher likelihood values. Parameter estimation involves determining the mean and uncertainty of the parameter distribution by examining the locations where the walker frequently visits and the extent of its deviations within the parameter space. To approximate the shape of the posterior distribution, the walker needs to take a sufficiently large number of steps. In Bayesian model selection, the evidence for each model is computed individually and then compared. This task usually demands substantial computational resources and often utilizes nested sampling methods. However, simpler tools like information criteria can offer reliable insights into model preferences when the posterior distributions closely resemble a Gaussian distribution.
Our study investigates Dark Energy Models within Finsler geometry and Riemannian geometry frameworks. We aim to determine the optimal parameter values for these models and estimate the present-day Hubble constant (\(H_0\)).
We employ Markov Chain Monte Carlo (MCMC) analysis using the Polychord package to explore the parameter space and obtain reliable estimates efficiently. This approach ensures robust exploration of the parameter space and accurate parameter estimation.
\subsection{Cosmological Data Sets}
\subsubsection{Cosmic Chronometers (CC)}
Cosmic chronometers are reliable probes for studying cosmic expansion, providing a model-independent method to measure the Hubble constant and the expansion rate. By analyzing the age and metallicity of nearby passive galaxies, we can estimate the expansion rate \( H_{\mathrm{CC}}(z) \) at a given redshift \( z_{\mathrm{CC}} \). This estimation is based on the observation that the expansion rate can be approximated by the ratio of the redshift difference to the time difference, adjusted by the redshift itself: \( H_{\mathrm{CC}}(z) \approx -(\Delta z_{\mathrm{CC}} / \Delta t) / (1+z_{\mathrm{CC}}) \). We have gathered cosmic chronometer data from various sources \cite{101,102,103,104} covering a redshift range of approximately \( 0.07 \lesssim z \lesssim 1.97 \). These data provide direct constraints on the universe's expansion history. To evaluate the consistency between theoretical predictions and cosmic chronometer measurements across different redshifts, we calculate the chi-squared distance \( \chi_{\mathrm{CC}}^{2} \):
\begin{equation}
\chi_{\mathrm{CC}}^{2} = \sum_{i=1}^{31} \left( \frac{H(z_{\mathrm{CC}}) - H_{\mathrm{CC}}(z_{\mathrm{CC}})}{\sigma_{\mathrm{CC}}(z_{\mathrm{CC}})} \right)^{2}.
\end{equation}
In this equation, \( H_{\mathrm{CC}}(z_{\mathrm{CC}}) \pm \sigma_{\mathrm{CC}}(z_{\mathrm{CC}}) \) represents the cosmic chronometer measurements of the expansion rate at redshift \( z_{\mathrm{CC}} \), along with their uncertainties. We assume that the cosmic chronometer observations are uncorrelated.
\subsubsection{Type Ia Supernova (SNIa)}
One of the most renowned and commonly utilized cosmological tools is the study of distant Type Ia Supernovae. These supernova explosions are exceptionally luminous, often rivaling the brightness of their host galaxies \cite{smith2020}. Their observed light curves typically exhibit peak brightness levels that remain relatively unaffected by distance, making them invaluable as standard candles. A key metric used to constrain cosmological models is the observed distance modulus, denoted as \(\mu_{\text{obs}}\). For our analysis, we use the most recent dataset available, the binned Pantheon dataset described in \cite{smith2020}. The likelihood function for the Type Ia Supernovae data, \(L_{\text{SNIa}}(Y;M)\), is expressed as:
\begin{equation}
L_{\text{SNIa}}(Y;M) \propto \exp \left( -\frac{1}{2} \sum_{i=1}^{40} m_i C_{\text{cov}}^{-1} \, \text{cov}_m^\dagger \, i \right).
\end{equation}
In this context, \(Y\) represents the vector of free parameters of the cosmological model. The term \(m_i\) is defined as the difference between the observed distance modulus \(\mu_{\text{obs},i}\) and the theoretical distance modulus \(\mu_{\text{theor}}(z_i)\), adjusted by \(M\). The theoretical distance modulus \(\mu_{\text{theor}}\) is calculated using the standard luminosity distance formula:
\begin{equation}
\mu_{\text{theor}} = 5 \log \left( \frac{D_L}{1 \, \text{Mpc}} \right) + 25,
\end{equation}
where \(D_L\) represents the luminosity distance, given by \(D_L = c (1 + z) \int_0^z \frac{dz'}{H(z')}\). This formula is applicable in a flat Friedmann-Robertson-Walker (FRW) space-time, irrespective of the underlying cosmology. Finally, \(C_{\text{cov}}\) denotes the covariance matrix associated with the binned Pantheon dataset. The parameter \(M\) is an intrinsic free parameter within the Pantheon dataset, capturing various observational uncertainties, including those related to host galaxy properties. To broaden our observational spectrum, we include 24 binned quasar distance modulus data from \cite{106}, a sample of 162 Gamma-Ray Bursts (GRBs) as described in \cite{107}
\subsubsection{Baryon Acoustic Oscillations (BAO)}
\begin{figure*}
   \begin{minipage}{0.49\textwidth}
     \centering
   \includegraphics[scale=0.55]{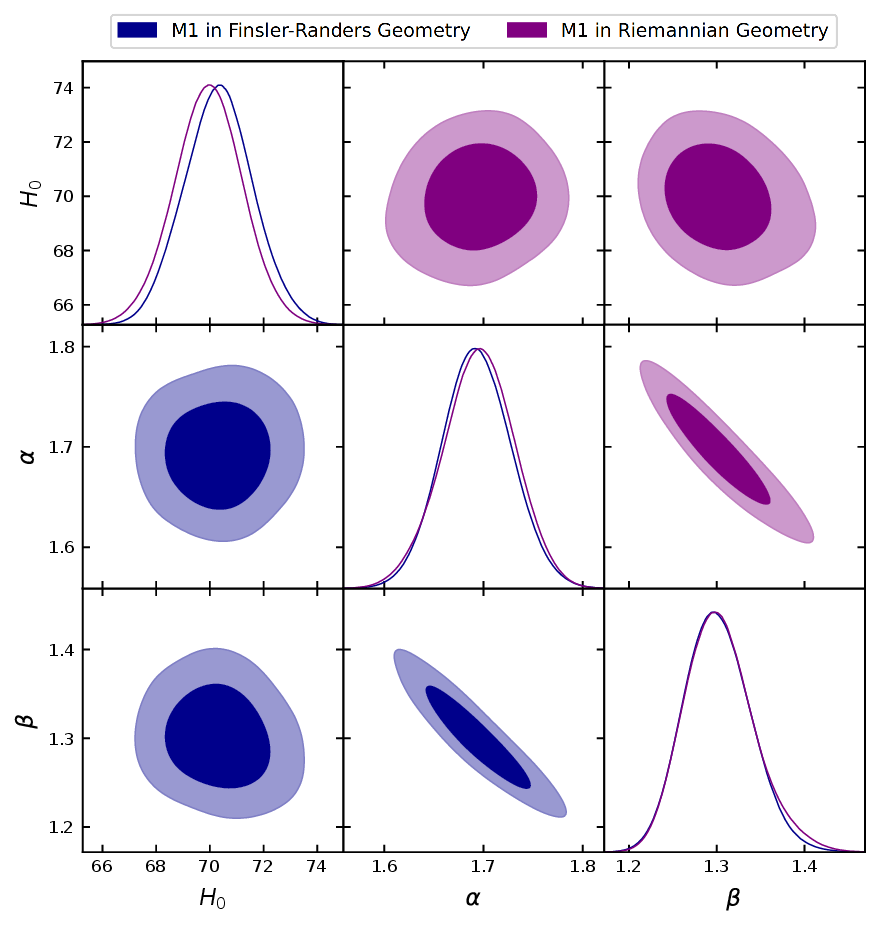}
\caption{MCMC confidence contours at 1$\protect\sigma$ and 2$\protect\sigma$ for Model 1.}\label{fig_1}
   \end{minipage}\hfill
   \begin{minipage}{0.49\textwidth}
     \centering
    \includegraphics[scale=0.55]{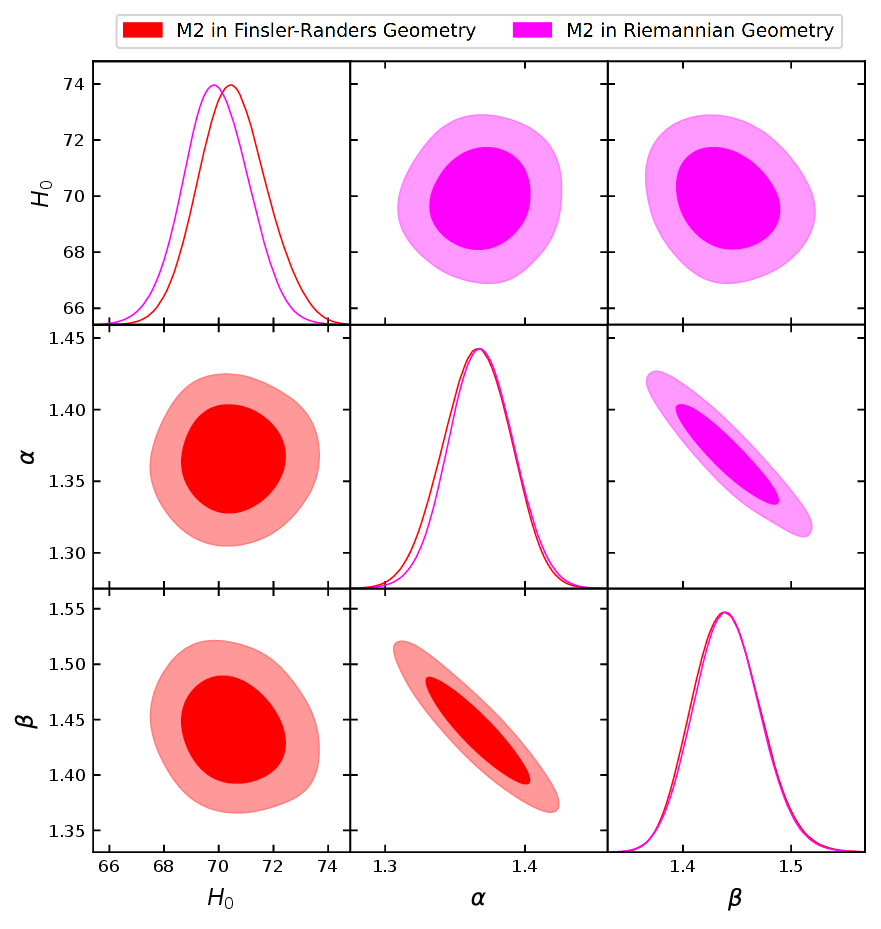}
\caption{MCMC confidence contours at 1$\protect\sigma$ and 2$\protect\sigma$ for Model 2.}\label{fig_2}
   \end{minipage}
\end{figure*}

For our analysis, we utilize a comprehensive list of BAO data points, including effective redshifts (\(z_{\text{eff}}\)), observables, measurements, and associated errors, as provided in \cite{Benisty}. The BAO feature in the transverse direction offers a measurement of \(\frac{D_H(z)}{r_d} = \frac{c}{H(z)r_d}\), where \(D_H(z)\) is the Hubble distance at redshift \(z\), and \(r_d\) is the sound horizon at the drag epoch. In a flat Universe, the comoving angular diameter distance is given by $D_M = \frac{c}{H_0} \int_0^z \frac{dz'}{E(z')},$. where \(E(z)\) is the dimensionless Hubble parameter. The angular diameter distance \(D_A\) is defined as \(D_A = \frac{D_M}{1 + z}\), and the volume-averaged distance \(D_V(z)/r_d\) contains information about the BAO peak coordinates. The volume-averaged distance is given by$D_V(z) = \left[ z D_H(z) D_M^2(z) \right]^{1/3},$. with \(r_d\) being the sound horizon at the drag epoch, measured as 147.1 Mpc \cite{Aghanim2020}. To compare theoretical predictions with observational BAO data, we calculate the distance function \(\chi_{\mathrm{BAO}}^{2}\) as follows:
\begin{equation}
\chi_{\mathrm{BAO}}^{2} = \sum_{i=1}^{17} \left( \frac{H(z_{\mathrm{BAO}}) - H_{\mathrm{BAO}}(z_{\mathrm{BAO}})}{\sigma_{\mathrm{BAO}}(z_{\mathrm{BAO}})} \right)^{2},
\end{equation}
where \(H_{\mathrm{BAO}}(z_{\mathrm{BAO}}) \pm \sigma_{\mathrm{BAO}}(z_{\mathrm{BAO}})\) represents the BAO-derived measurements of the expansion rate at redshift \(z_{\mathrm{BAO}}\), scaled by the sound horizon radius. In our analysis, we have employed a nested sampler, implemented within the open-source package Polychord \cite{108}, in conjunction with the GetDist package \cite{109} for presenting the results. Figs \ref{fig_1} and \ref{fig_2} depict the \(68\%\) and \(95\%\) confidence levels for key cosmological parameters in both the Riemannian and Finslerian dark energy models. Table \ref{mcmcresults} presents the best-fit values of the model parameters $\alpha$ and $\beta$, along with the current Hubble constant $H_{0}$.
\begin{table*}
\begin{center}
\setlength{\arrayrulewidth}{1.6pt}
\begin{tabular}{|c|c|c|c|c|c|}
\hline
\multicolumn{6}{|c|}{MCMC Results} \\ \hline\hline
Finslerian Geometry & Parameters & Best-fit Value & Riemannian Geometry & Parameters & Best-fit Value  \\ \hline
\multirow{8}{*}& $\makecell{H_0\\~~~~~~~~~~~~~~~~}$ & $70.326621_{\pm 1.381923}^{\pm 2.380077}$ & \multirow{8}{*} & $H_0$ & $69.946236_{\pm 1.203929}^{\pm 2.522770}$\\
Model 1 &$\makecell{\alpha\\~~~~~~~~~~~~~~~~}$ & $1.693215_{\pm 0.033153}^{\pm 0.066342}$ & Model 1 & $\alpha$ & $1.695306_{\pm 0.033177}^{\pm 0.073840}$ \\
& $\makecell{\beta\\~~~~~~~~~~~~~~~~}$ & $1.300786_{\pm 0.037316}^{\pm 0.068124}$ & & $\beta$ & $1.302990_{\pm 0.040574}^{\pm 0.069706}$\\
\hline\hline
\multirow{8}{*}& $\makecell{H_0\\~~~~~~~~~~~~~~~~}$ & $70.517486_{\pm 1.219293}^{\pm 2.347270}$ & \multirow{8}{*} & $H_0$ & $69.901832_{\pm 1.073915}^{\pm 2.333254}$\\
Model 2 & $\makecell{\alpha\\~~~~~~~~~~~~~~~~}$ & $1.365617_{\pm 0.024308}^{\pm 0.046564}$ & Model 2 & $\alpha$ & $1.368668_{\pm 0.022250}^{\pm 0.043644}$ \\
& $\makecell{\beta\\~~~~~~~~~~~~~~~~}$ & $1.440327_{\pm 0.031578}^{\pm 0.055222}$ & & $\beta$ & $1.440785_{\pm 0.029173}^{\pm 0.058645}$\\
\hline
\end{tabular}
\caption{Summary of MCMC Results.}\label{mcmcresults}
\end{center}
\end{table*}
\section{Observational and Theoretical comparisons of the Hubble Parameter, Hubble Difference, and Apparent Magnitude}\label{sec7}
After obtaining the best-fit values for both dark energy parametrizations in Riemannian and Finslerian models, it is necessary to compare these models with real observational datasets and the standard $\Lambda$CDM paradigm. To conduct a comprehensive comparative study, we will plot the Hubble function for all three scenarios: the Riemannian model, the Finslerian model, and the $\Lambda$CDM model. We will use the 31 Cosmic Chronometer (CC) datasets using their corresponding error bars for this analysis. Additionally, we will plot the difference between the Hubble functions of the Riemannian and Finslerian models relative to the $\Lambda$CDM model, denoted as \( H_{\text{Model}}(z) - H_{\Lambda\text{CDM}}(z) \), against the CC datasets. Furthermore, we will plot the apparent magnitude in each mode and compare these with each other and with actual observational data. This comparison is essential for assessing the viability and accuracy of both dark energy parametrizations in explaining the Universe's accelerated expansion. Analyzing the Hubble function allows us to directly examine the rate of expansion over time while comparing \( H_{\text{Model}}(z) - H_{\Lambda\text{CDM}}(z) \) helps us understand deviations from the standard model. The apparent magnitude comparison provides insights into the models' predictions of luminosity distance, which is crucial for understanding cosmological distances and the scale of the Universe.\\\\
\subsection{Hubble Parameter $H(z)$}
\begin{figure*}[htp]
   \begin{minipage}{0.49\textwidth}
     \centering
   \includegraphics[scale=0.4]{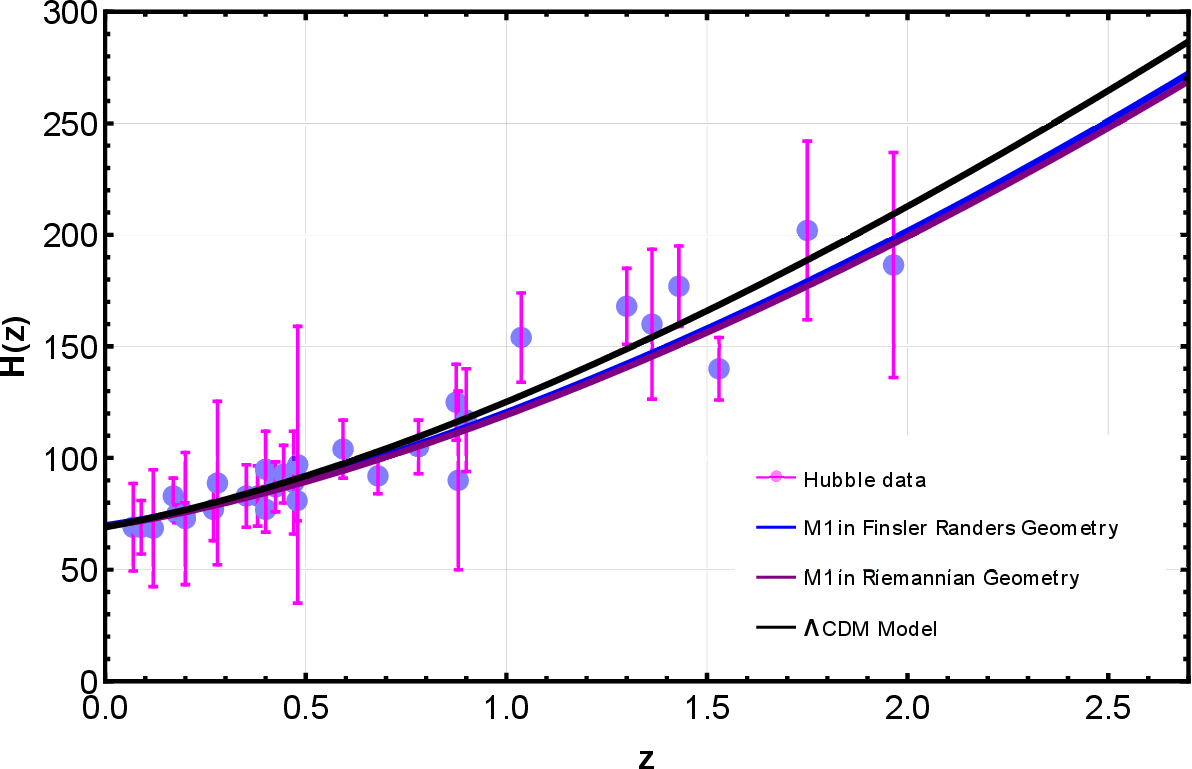}
\caption{Evolution of the Hubble Parameter as a function of redshift in Riemannian and Finslerian Modes}\label{fig_3}
   \end{minipage}\hfill
   \begin{minipage}{0.49\textwidth}
     \centering
    \includegraphics[scale=0.4]{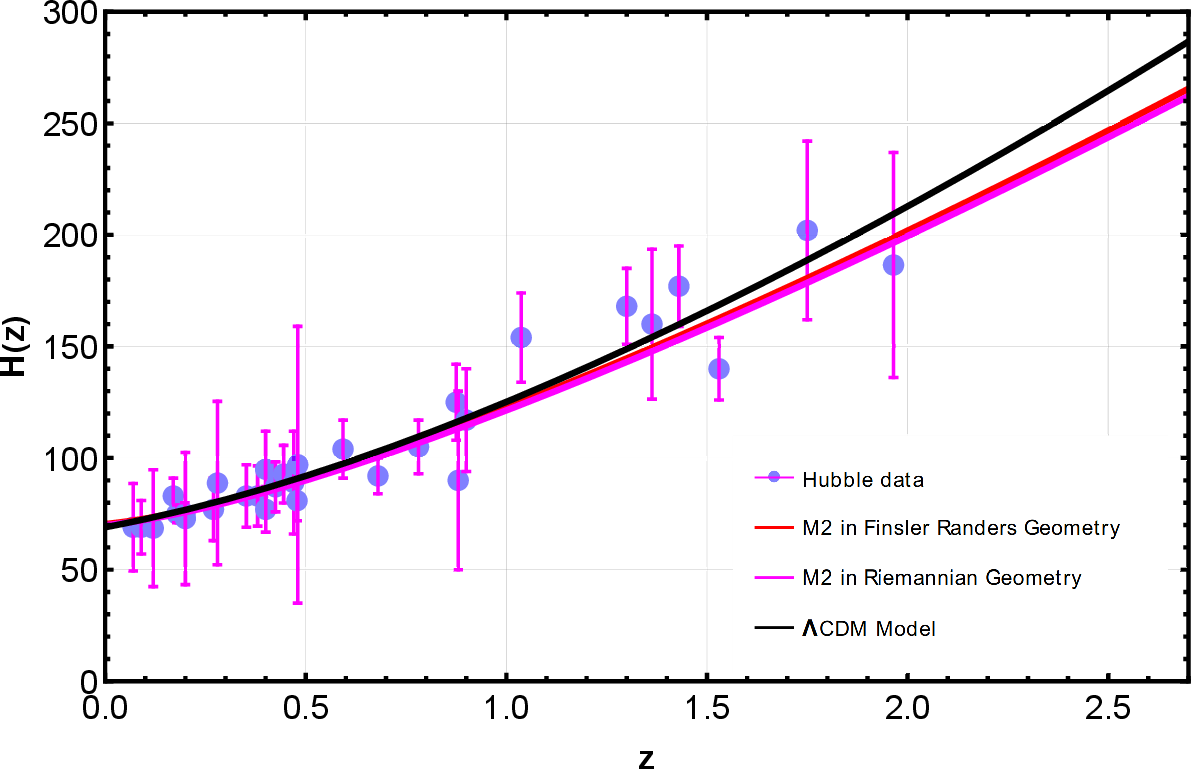}
\caption{Evolution of the Hubble Parameter as a function of redshift in Riemannian and Finslerian Modes}\label{fig_4}
   \end{minipage}
\end{figure*}

Figs.~\ref{fig_3} and \ref{fig_4} illustrate the evolution of the Hubble parameters as a function of redshift for both the Riemannian and Finslerian models, with the $\Lambda$CDM model (black line) included for comparison, using cosmological parameters $\Omega_{\mathrm{m0}} = 0.3$ and $\Omega_\Lambda = 0.7$. In Fig.~\ref{fig_3} for model M1, a deviation between the Riemannian (purple line) and Finslerian (blue line) models and the $\Lambda$CDM model is evident at \(z > 0.6\). However,  all three models exhibit close agreement at \(z < 0.6\). Similarly, in Fig.~\ref{fig_4} for model M2, a deviation between the Riemannian (magenta line) and Finslerian (red line) models and the $\Lambda$CDM model is observed at \(z > 1.2\) and close agreement at \(z < 1.2\).
\subsection{Hubble Difference \( \Delta H(z) \)}
\begin{figure*}[htp]
   \begin{minipage}{0.49\textwidth}
     \centering
   \includegraphics[scale=0.4]{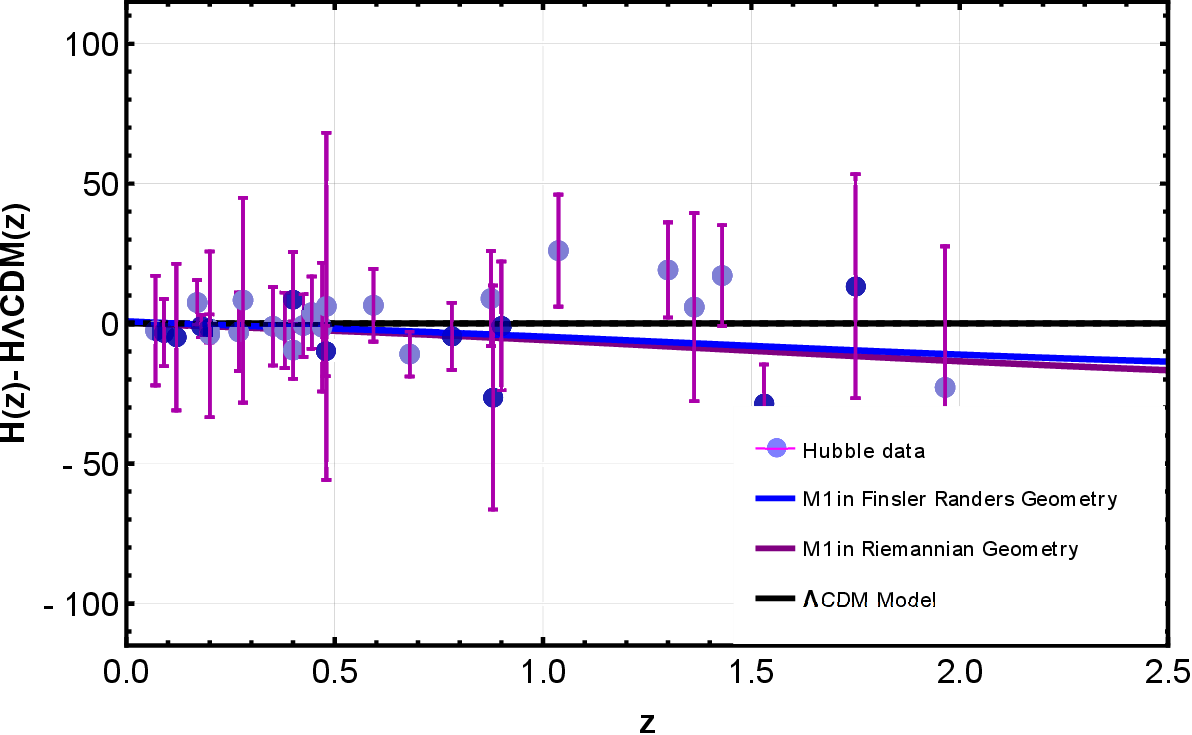}
\caption{Evolution of the Hubble Difference in Riemannian and Finslerian Models}\label{fig_5}
   \end{minipage}\hfill
   \begin{minipage}{0.49\textwidth}
     \centering
    \includegraphics[scale=0.4]{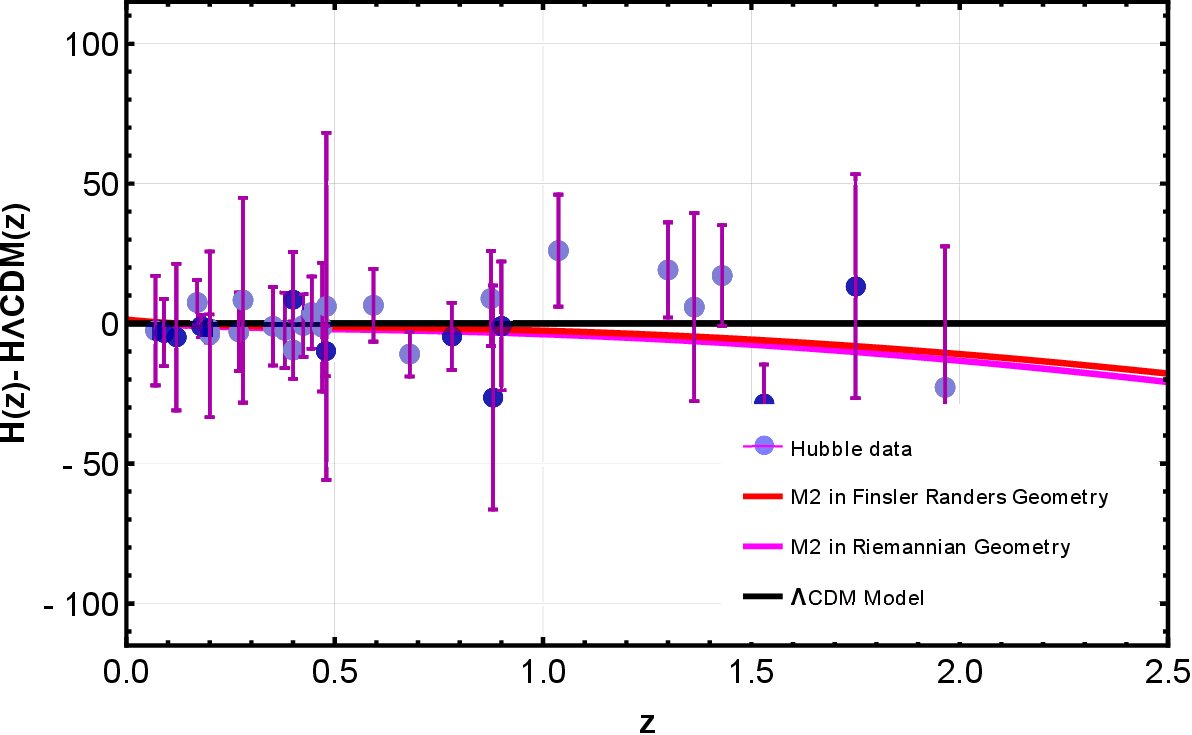}
\caption{Evolution of the Hubble Difference in Riemannian and Finslerian Models.}\label{fig_6}
   \end{minipage}
\end{figure*}

Figs.~\ref{fig_5} and \ref{fig_6} illustrate the evolution of the difference between the Hubble parameter \( H \) predicted by specific cosmological models, Riemannian and Finslerian in our case and the Hubble parameter \( H \) predicted by the $\Lambda$CDM model, both evaluated at a given redshift \( z \). Fig.~\ref{fig_5} shows \( \Delta H(z) = H_{\text{M1}}(z) - H_{\Lambda\text{CDM}}(z) \). It can be observed that \( \Delta H(z) \) is significant at \( z > 0.6 \) for both the Riemannian (purple line) and Finslerian (blue line) models compared to the $\Lambda$CDM model. However, at \( z < 0.6 \), \( \Delta H(z) \) becomes negligible. Fig.~\ref{fig_6} for model M2 shows \( \Delta H(z) \). A significant difference between the Riemannian (magenta line) and Finslerian (red line) models and the $\Lambda$CDM model is observed at \( z > 1 \). However, \( \Delta H(z) \) becomes negligible at \( z < 1 \).\\\\
\subsection{Distance Modulus \(\mu(z)\)}
\begin{figure*}[htp]
   \begin{minipage}{0.49\textwidth}
     \centering
   \includegraphics[scale=0.45]{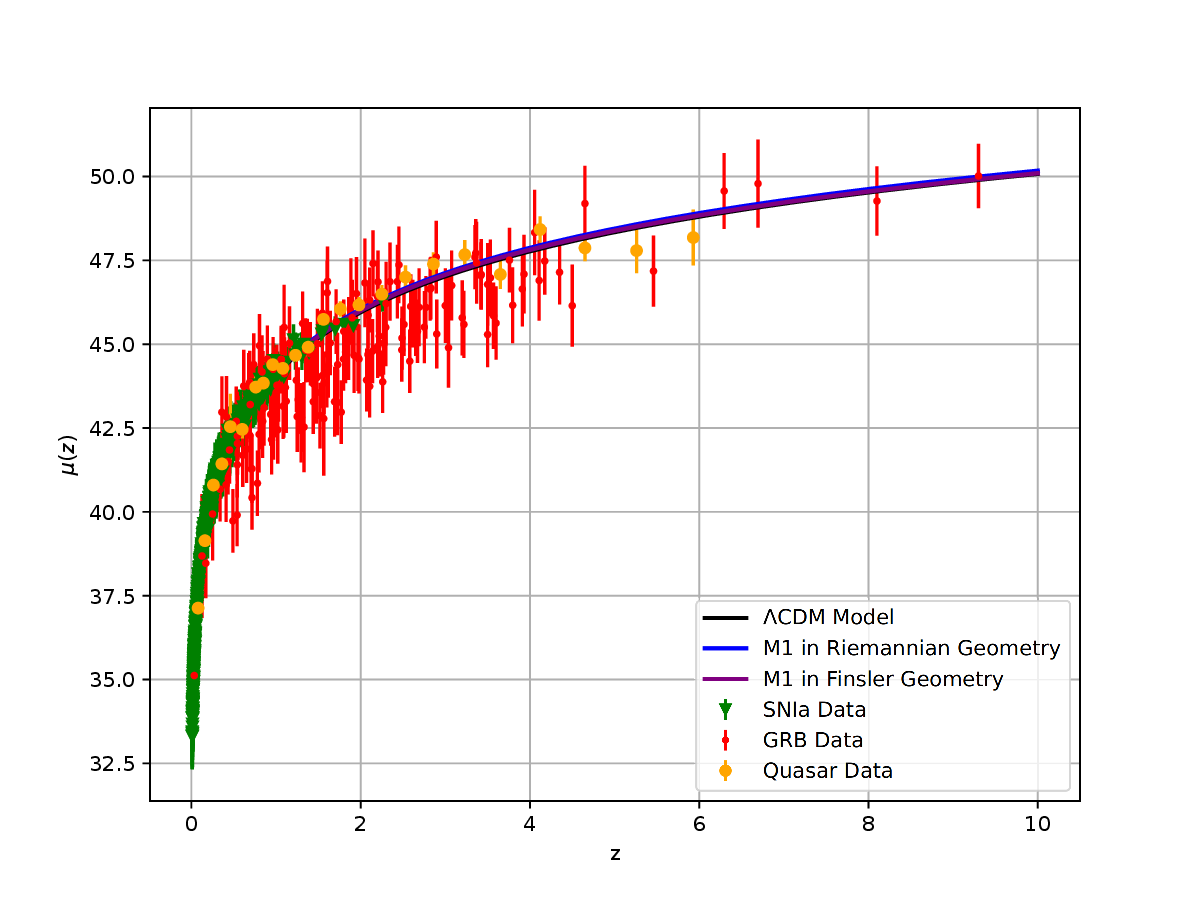}
\caption{The evolution of Distance Modulus in Riemannian and Finslerian modes.}\label{fig_7}
   \end{minipage}\hfill
   \begin{minipage}{0.49\textwidth}
     \centering
    \includegraphics[scale=0.43]{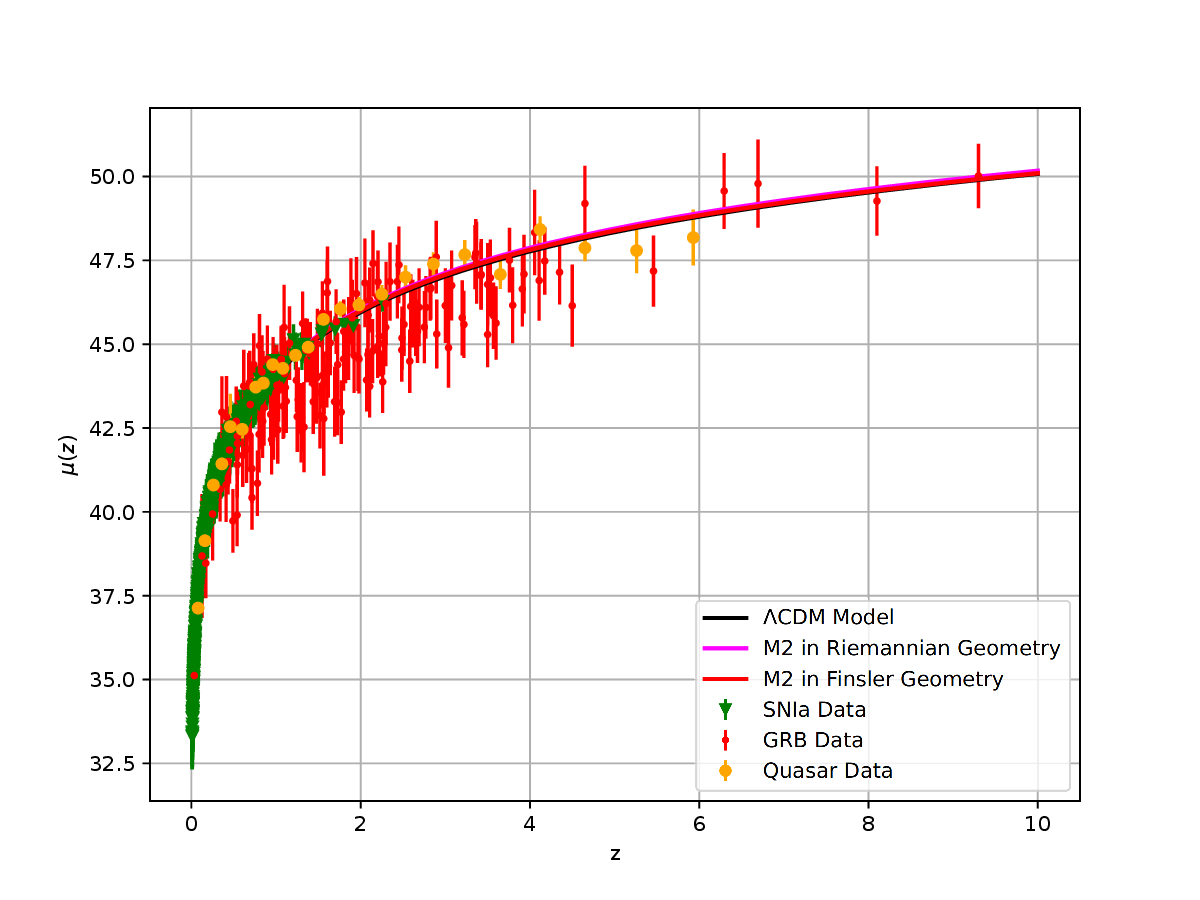}
\caption{The evolution of Distance Modulus in Riemannian and Finslerian modes.}\label{fig_8}
   \end{minipage}
\end{figure*}

Figs.~\ref{fig_7} and \ref{fig_8} depict the evolution of the distance modulus \(\mu(z)\) for a Type Ia supernova (SINa), Gamma Ray Burst (GRB), and Quasar (Q). The distance modulus is an astronomical measure used to quantify the apparent brightness of celestial objects based on their redshift \(z\), which indicates how much the Universe has expanded since the object's light was emitted. It is related to the luminosity distance \(D_{L}\) by the formula: \(\mu = 5 \log_{10}(D_L) + 25\), where \( \mu \) is the distance modulus and \( D_L \) is the luminosity distance in megaparsecs (Mpc). In the standard cosmological model, the luminosity distance is calculated as: \( D_L = (1 + z) \cdot D_C(z) \). Here, \( z \) is the redshift, and \( D_C(z) \) is the comoving distance, which represents the spatial distance between the observer and the object, accounting for the expansion of the Universe. The comoving distance \( D_C(z) \) is given by an integral involving the inverse of the Hubble parameter \( H(z) \): \( D_C(z) = \int_0^z \frac{c \, dz'}{H(z')} \), where \( c \) is the speed of light and \( H(z) \) is the Hubble parameter at redshift \( z \). Fig.~\ref{fig_7} illustrates a close agreement between the Riemannian (purple line) and Finslerian (blue line) modes with $\Lambda$CDM model and SNIa dataset at low redshift. However,  differences become evident at high redshifts when considering the SNIa dataset. Fig.~\ref{fig_8} exhibits a similar trend in model M2, where both Riemannian (magenta line) and Finslerian (red line) modes closely resemble the $\Lambda$CDM model.\\\\
\section{Cosmographic Parameter}
\subsection{Deceleration Parameter}
\begin{figure*}[htp]
   \begin{minipage}{0.49\textwidth}
     \centering
   \includegraphics[scale=0.5]{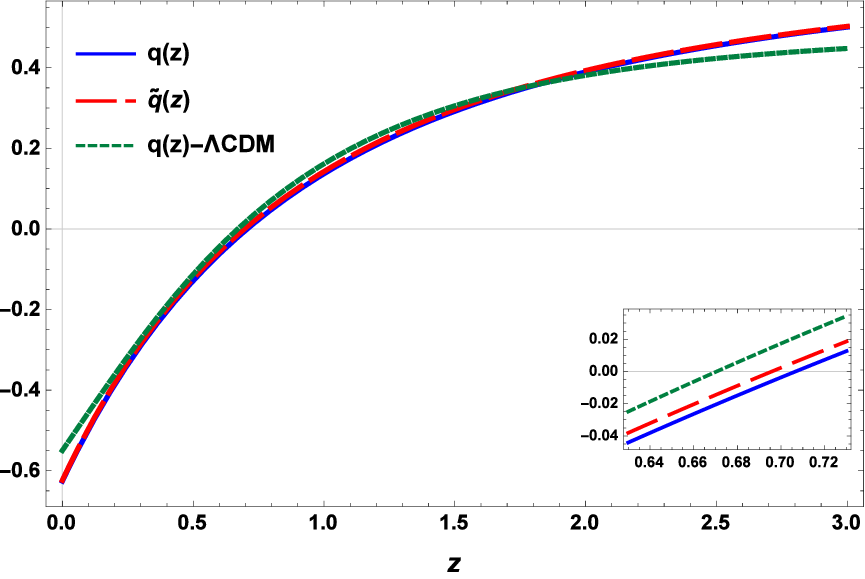}
\caption{The evolution of deceleration parameter from past $(z = 4)$ to far future with a phase transition
in Riemannian and Finslerian modes.}\label{fig_9}
   \end{minipage}\hfill
   \begin{minipage}{0.49\textwidth}
     \centering
    \includegraphics[scale=0.5]{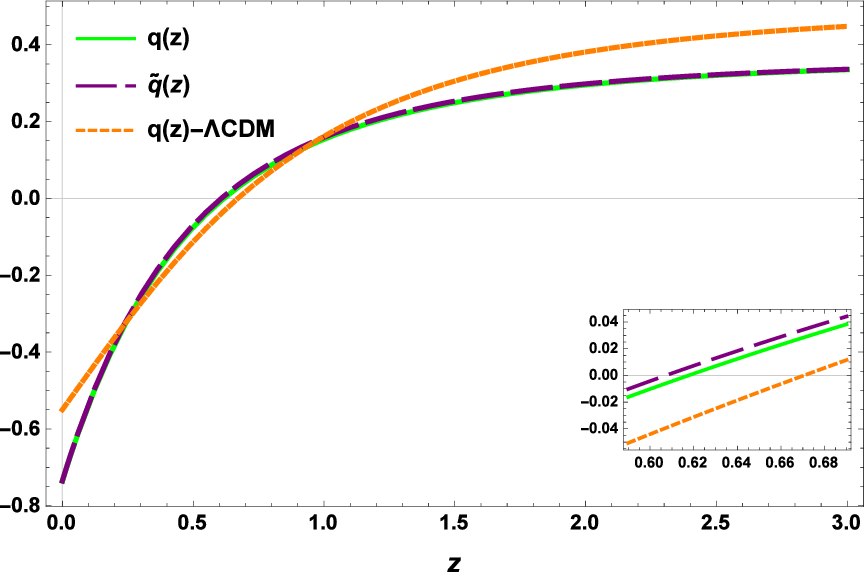}
\caption{The evolution of deceleration parameter from past $(z = 4)$ to far future with a phase transition
in Riemannian and Finslerian modes.}\label{fig_10}
   \end{minipage}
\end{figure*}

The deceleration parameter (DP), \(q\), is a dimensionless measure in cosmology that describes the rate of change of the universe's expansion. It is defined as \(q = -\frac{\ddot{a}a}{\dot{a}^2}\), where \(a\) is the scale factor of the universe, \(\dot{a}\) is its first derivative with respect to time (the expansion rate), and \(\ddot{a}\) is its second derivative (the acceleration of the expansion). A positive \(q\) indicates a decelerating expansion, while a negative \(q\) indicates an accelerating expansion. Understanding \(q\) is crucial for studying the dynamics of the universe, including the transition from a decelerated to an accelerated phase, which is associated with the influence of dark energy. In the case of Finslerian Geometry, the expression of the deceleration parameter in terms of redshift \(z\) for Models 1 and 2 is provided in Table \ref{tab_5}. Figs~\ref{fig_9} and \ref{fig_10} show the evolution of the deceleration parameter for Models 1 and 2 in both Riemannian and Finslerian geometries, compared with the \(\Lambda\)CDM model.
\begin{table*}[htp]
\centering
\begin{tabular}{|l |c|c|}
\hline
Models & $t(z)$ & $\tilde{q}(z)$ \\
\hline
   \text{Model 1} & $\frac{k_1}{1+\left[\beta(1+z)\right]^\alpha}$ & $\frac{\frac{\left(1+\beta^\alpha\right)^2(1+z)^\alpha}{H_0\left(1+\left[\beta(1+z)\right]^\alpha\right)^2}
   \rm{e}^{-\frac{\left(1+\beta^\alpha\right)^2}{\alpha H_0\beta^\alpha\left(1+\left[\beta(1+z)\right]^\alpha\right)}}-1+\alpha-2\alpha\left(1+\left[\beta(1+z)\right]^\alpha\right)^{-1}}
   {\frac{-\left(1+\beta^\alpha\right)^2(1+z)^\alpha}{H_0\left(1+\left[\beta(1+z)\right]^\alpha\right)^2}
   \rm{e}^{-\frac{\left(1+\beta^\alpha\right)^2}{\alpha H_0\beta^\alpha\left(1+\left[\beta(1+z)\right]^\alpha\right)}}+1}$\\
\hline
\text{Model 2} & $\frac{\sqrt{k_1}}{\sqrt{1+\left[\beta(1+z)\right]^{2\alpha}}}$ &$\frac{\frac{\left(1+\beta^{2\alpha}\right)^\frac{3}{2}(1+z)^{2\alpha}}{H_0\left(1+\left[\beta(1+z)\right]^{2\alpha}\right)^\frac{3}{2}}
   \rm{e}^{-\frac{\left(1+\beta^{2\alpha}\right)^\frac{3}{2}}{\alpha H_0\beta^{2\alpha}\left(1+\left[\beta(1+z)\right]^{2\alpha}\right)^\frac{1}{2}}}-1+\alpha-
   \frac{3\alpha}{1+\left[\beta(1+z)\right]^{2\alpha}}}{-\frac{\left(1+\beta^{2\alpha}\right)^\frac{3}{2}(1+z)^{2\alpha}}
   {H_0\left(1+\left[\beta(1+z)\right]^{2\alpha}\right)^\frac{3}{2}}
   \rm{e}^{-\frac{\left(1+\beta^{2\alpha}\right)^\frac{3}{2}}{\alpha H_0\beta^{2\alpha}\left(1+\left[\beta(1+z)\right]^{2\alpha}\right)^\frac{1}{2}}}+1}$ \\
\hline
\end{tabular}
\caption{Expression of the deceleration parameter}\label{tab_5}
\end{table*}
\subsection{Jerk and Snap Parameters}
As we know, \(q(t)\) represents the deceleration parameter, which measures the Universe's acceleration. Now, we are going to introduce other parameters such as \(j(t) =\frac{\dddot{a}}{aH^3}\) (jerk parameter), and \(s(t) =\frac{a^{(4)}}{aH^4}\) (snap parameter). These parameters play significant roles in the cosmographic analysis of the Universe, specifically in understanding the behavior of acceleration, jerk, and snap, and help distinguish between different dark energy models. Table \ref{tab_8} displays the mathematical expressions of the jerk parameters in terms of redshift.
\begin{itemize}
  \item Anisotropic jerk parameter: $\tilde{j}(t)=\frac{\dddot{a}+\frac{3}{2}\ddot{a}\dot{\mathfrak{u}}_0}{a\tilde{H}^3}$,\\
  \item Anisotropic snap parameter: $\tilde{s}(t)=\frac{a^{(4)}+2\dddot{a}\dot{\mathfrak{u}}_0}{a\tilde{H}^4}$,\\
\end{itemize}
\begin{table*}[htbp]
\centering
\begin{tabular}{|l |c|}
\hline
Models  & $\tilde{j}(z)$    \\
\hline
   \text{Model 1} & $\frac{-3\frac{\left(1+\beta^\alpha\right)^2(1+z)^\alpha}{H_0\left(1+\left[\beta(1+z)\right]^\alpha\right)^2}
   \rm{e}^{-\frac{\left(1+\beta^\alpha\right)^2}{\alpha H_0\beta^\alpha\left(1+\left[\beta(1+z)\right]^\alpha\right)}}\left(\frac{\alpha}{1+\beta^\alpha(1+z)^\alpha}
   -\frac{\alpha}{2}+\frac{1}{2}\right)+1+\alpha(2\alpha-3)+\frac{6\alpha}{1+\beta^\alpha(1+z)^\alpha}\left[1-\alpha+
   \frac{\alpha}{1+\beta^\alpha(1+z)^\alpha}\right]}{-\frac{3}{2}\frac{\left(1+\beta^\alpha\right)^2(1+z)^\alpha}
   {H_0\left(1+\left[\beta(1+z)\right]^\alpha\right)^2}
   \rm{e}^{-\frac{\left(1+\beta^\alpha\right)^2}{\alpha H_0\beta^\alpha\left(1+\left[\beta(1+z)\right]^\alpha\right)}}+1}$  \\
   \hline
\text{Model 2} & $\frac{\frac{-9}{2}\frac{\left(1+\beta^{2\alpha}\right)^\frac{3}{2}(1+z)^{2\alpha}}{H_0\left(1+\left[\beta(1+z)\right]^{2\alpha}\right)^\frac{3}{2}}
   \rm{e}^{-\frac{\left(1+\beta^{2\alpha}\right)^\frac{3}{2}}{\alpha H_0\beta^{2\alpha}\left(1+\left[\beta(1+z)\right]^{2\alpha}\right)^\frac{1}{2}}}\left(\frac{\alpha}{1+\left[\beta(1+z)\right]^{2\alpha}}-\frac{\alpha}{3}+\frac{1}{3}\right)+1-3\alpha+2\alpha^2+
   \frac{12\alpha^2}{\left(1+\left[\beta(1+z)\right]^{2\alpha}\right)^2}+\frac{3\alpha(3-2\alpha)}{1+\left[\beta(1+z)\right]^{2\alpha}}}{-\frac{3}{2}\frac{\left(1+\beta^{2\alpha}\right)^\frac{3}{2}(1+z)^{2\alpha}}
   {H_0\left(1+\left[\beta(1+z)\right]^{2\alpha}\right)^\frac{3}{2}}
   \rm{e}^{-\frac{\left(1+\beta^{2\alpha}\right)^\frac{3}{2}}{\alpha H_0\beta^{2\alpha}\left(1+\left[\beta(1+z)\right]^{2\alpha}\right)^\frac{1}{2}}}+1}$ \\
\hline
\end{tabular}
\caption{Expression of the jerk parameter.}\label{tab_8}
\end{table*}
\begin{figure*}[htp]
   \begin{minipage}{0.49\textwidth}
     \centering
   \includegraphics[scale=0.5]{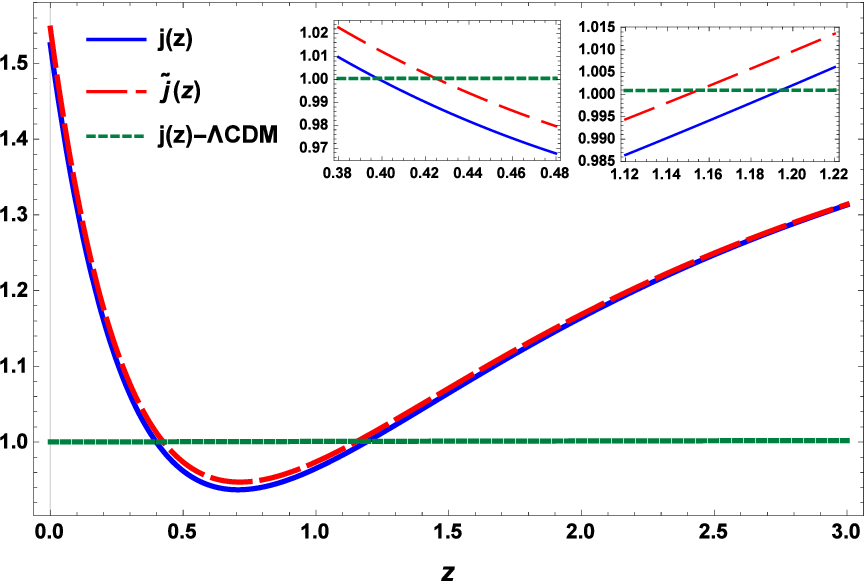}
\caption{Jerk parameter vs. $z$ in Riemannian and Finslerian modes.}\label{fig_11}
   \end{minipage}\hfill
   \begin{minipage}{0.49\textwidth}
     \centering
    \includegraphics[scale=0.5]{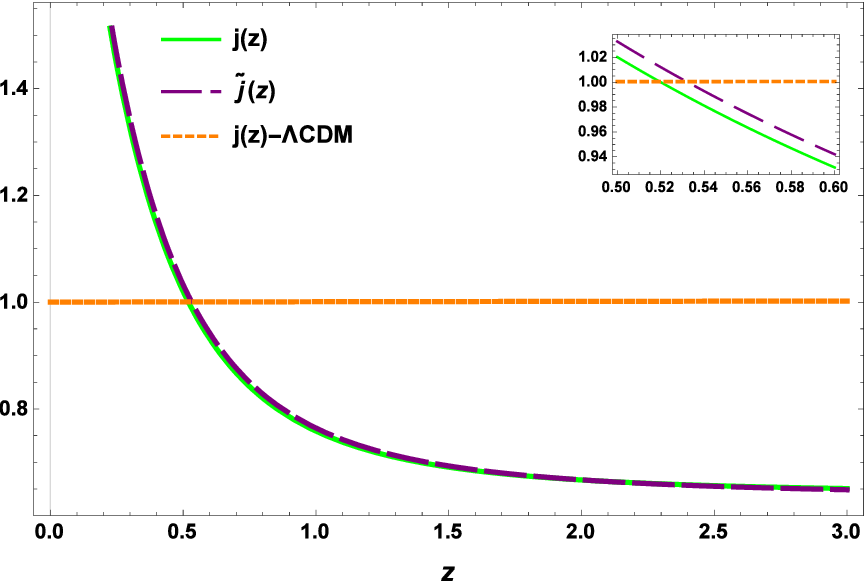}
\caption{Jerk parameter vs. $z$ in Riemannian and Finslerian modes.}\label{fig_12}
   \end{minipage}
\end{figure*}
\begin{figure*}[htp]
   \begin{minipage}{0.49\textwidth}
     \centering
   \includegraphics[scale=0.5]{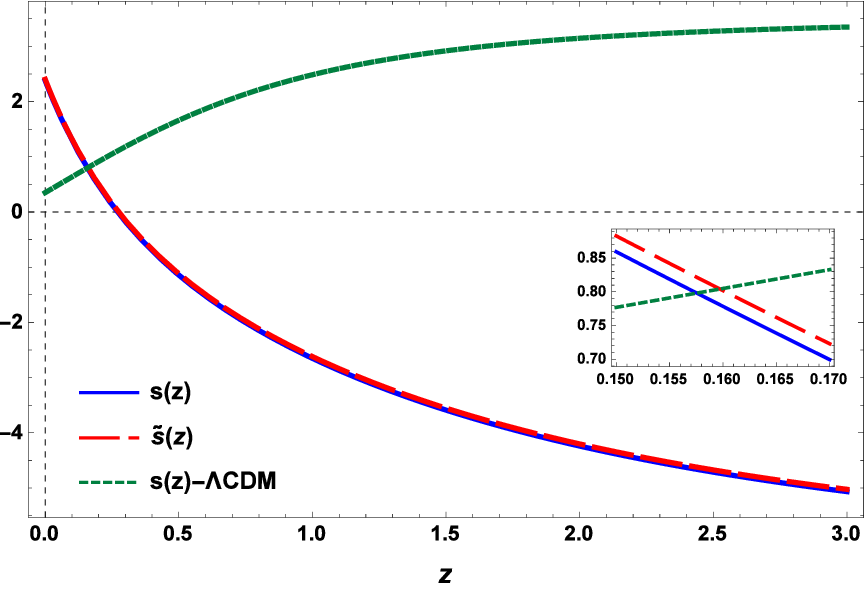}
\caption{Snap parameter vs. $z$ in Riemannian and Finslerian modes.}\label{fig_13}
   \end{minipage}\hfill
   \begin{minipage}{0.49\textwidth}
     \centering
    \includegraphics[scale=0.5]{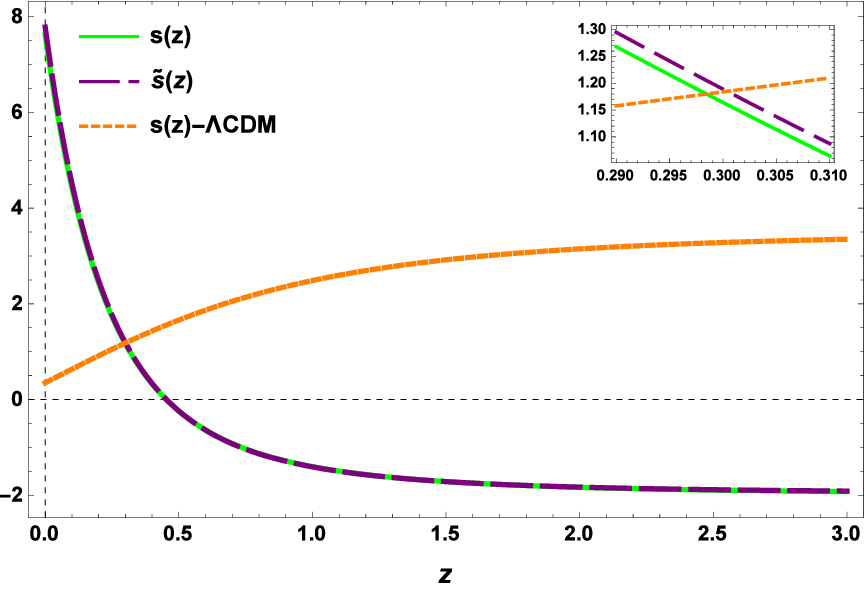}
\caption{Snap parameter vs. $z$ in Riemannian and Finslerian modes.}\label{fig_14}
   \end{minipage}
\end{figure*}
\section{Statefinder Diagnostic}
Statefinder diagnostics \cite{Sahni, Alam, Sami, Myrzakulov} is generally utilized to denote various dark energy models and compare their behavior using the higher-order derivatives of the scale factor. The parameters are $\mathfrak{s}$ and $r$ are computed by using the following relations:
\begin{equation*}
r=j=\frac{\dddot{a}}{aH^3}, ~~~~ \mathfrak{s}=\frac{r-1}{3\left(q-\frac{1}{2}\right)}.
\end{equation*}
The statefinder diagnostics pairs are made as {s, r} and {q, r} wherein various trajectories in the $\mathfrak{s}-r$ and $q-r$ planes scheme to see the temporal evolutions of various dark energy models.
The fixed points in this context are generally considered as $\{\mathfrak{s}, r\} = \{0, 1\}$ for the $\Lambda$CDM model and $\{\mathfrak{s}, r\} = \{1, 1\}$ for the SCDM (standard cold dark matter) model in FLRW background and the departures of any dark energy model from these fixed points are examined. The other diagnostic pair is $\{q, r\}$ and the fixed points considered are $\{q, r\}=\{-1, 1\}$ for the $\Lambda$CDM model
and $\{q, r\}=\{0.5, 1\}$ for SCDM model.The statefinder parameters for the considered Model 1 and Model 2 are computed as follows\\

\textbf{Model 1}\\
\begin{equation}
\tilde{r}=\tilde{j}, ~~~~ \tilde{\mathfrak{s}}=\frac{8\alpha\triangle_{1}}{27\triangle_{2}},
\end{equation}
where

\begin{eqnarray*}
\triangle_{1}&=&-\bigg[\left( \frac{3}{2}\,\alpha+{\frac{27}{8}} \right)  \left( \beta\, \left( 1+z
 \right)  \right) ^{2\,\alpha}+ \left(\frac{5}{4}\,\alpha+{\frac{15}{8}}
 \right)  \left( \beta\, \left( 1+z \right)  \right) ^{3\,\alpha}+
 \left( \frac{5}{4}\,\alpha-{\frac{15}{8}} \right)  \left( \beta\, \left( 1+z
 \right)  \right) ^{4\,\alpha}\\&+& \left( \frac{5}{4}\,\alpha-{\frac{27}{8}}
 \right)  \left( \beta\, \left( 1+z \right)  \right) ^{5\,\alpha}+
 \left( \alpha-{\frac{15}{8}} \right)  \left( \beta\, \left( 1+z
 \right)  \right) ^{6\,\alpha}+ \left( \frac{\alpha}{4}-\frac{3}{8} \right)  \left(
\beta\, \left( 1+z \right)  \right) ^{7\,\alpha}\\&+& \left( \alpha+{\frac
{15}{8}} \right)  \left( \beta\, \left( 1+z \right)  \right) ^{\alpha}
+\frac{\alpha}{4}+\frac{3}{8}\bigg]H_{0}^{2}{{\rm e}^{{\frac {2\,{\beta}^{-\alpha}+4+2\,{\beta}^{\alpha}}{\alpha\,
{H_0}\, \left( 1+ \left( \beta\, \left( 1+z \right)  \right) ^{\alpha} \right) }}}}+\bigg[\left( \frac{\alpha}{2}+\frac{9}{4} \right)  \left( \beta\, \left( 1+z \right)\right) ^{2\,\alpha}\\&+& \left( \frac{\alpha}{2}-\frac{9}{4} \right)  \left( \beta\,
 \left( 1+z \right)  \right) ^{3\,\alpha}+ \left( \alpha-{\frac{27}{8}
} \right)  \left( \beta\, \left( 1+z \right)  \right) ^{4\,\alpha}+
 \left( \frac{\alpha}{2}-{\frac{9}{8}} \right)  \left( \beta\, \left( 1+z
 \right)  \right) ^{5\,\alpha}\\&+& \left( \alpha+{\frac{27}{8}} \right)
 \left( \beta\, \left( 1+z \right)  \right) ^{\alpha}+\frac{\alpha}{2}+{\frac{9}{8}}\bigg]\left( 1+z \right) ^{\alpha}{H_0}\, \left( \frac{1}{2}\,{\beta}^{2\,\alpha}+{\beta}^{\alpha}+\frac{1}{2} \right) {{\rm e}^{{\frac {{\beta}^{-\alpha}+2+{
\beta}^{\alpha}}{\alpha\,{H_0}\, \left( 1+ \left( \beta\, \left( 1+
z \right)  \right) ^{\alpha} \right) }}}}\\&-&\frac{3}{4}\, \left( 1+z \right) ^{2\,\alpha} \bigg( {\beta}^{\alpha}+\frac{3}{2}\,{
\beta}^{2\,\alpha}+{\beta}^{3\,\alpha}+\frac{1}{4}\,{\beta}^{4\,\alpha}+\frac{1}{4}
 \bigg)  \bigg(  \left( \beta\, \left( 1+z \right)  \right) ^{\alpha}
- \left( \beta\, \left( 1+z \right)  \right) ^{2\,\alpha}\\&-& \left(
\beta\, \left( 1+z \right)  \right) ^{3\,\alpha}+1 \bigg),
\end{eqnarray*}

\begin{eqnarray*}
\triangle_{2}&=&\bigg[-{H_0}\, \left(  \left( \frac{-\alpha}{3}+\frac{1}{2} \right)  \left( \beta\,
 \left( 1+z \right)  \right) ^{2\,\alpha}+ \left( \beta\, \left( 1+z
 \right)  \right) ^{\alpha}+\frac{\alpha}{3}+\frac{1}{2} \right) {{\rm e}^{{\frac {{
\beta}^{-\alpha}+2+{\beta}^{\alpha}}{\alpha\,{H_0}\, \left( 1+
 \left( \beta\, \left( 1+z \right)  \right) ^{\alpha} \right) }}}}\\&+&
 \left( 1+z \right) ^{\alpha} \left( \frac{1}{2}\,{\beta}^{2\,\alpha}+{\beta}^
{\alpha}+\frac{1}{2} \right)\bigg]\bigg[-\frac{2}{3}\,{H_0}\, \left(  \left( \beta\, \left( 1+z \right)  \right) ^{
\alpha}+\frac{1}{2}\, \left( \beta\, \left( 1+z \right)  \right) ^{2\,\alpha}+
\frac{1}{2} \right)\\&.& {{\rm e}^{{\frac {{\beta}^{-\alpha}+2+{\beta}^{\alpha}}{
\alpha\,{H_0}\, \left( 1+ \left( \beta\, \left( 1+z \right)
 \right) ^{\alpha} \right) }}}}+ \left( 1+z \right) ^{\alpha} \left(\frac{1}{2}\,{\beta}^{2\,\alpha}+{\beta}^{\alpha}+\frac{1}{2} \right)\bigg]\left( 1+ \left( \beta\, \left( 1+z \right)  \right) ^{\alpha}
 \right) ^{3}.
\end{eqnarray*}
\textbf{Model 2}\\
\begin{equation}
\tilde{r}=\tilde{j}, ~~~~ \tilde{\mathfrak{s}}=\frac{-3\alpha\triangle_{3}}{2\triangle_{4}},
\end{equation}
\begin{eqnarray*}
\triangle_{3}&=&-\frac {28\,{{H_0}}^{2} \left( 1+ \left( \beta\, \left( 1+z
 \right)  \right) ^{2\,\alpha} \right) ^{\frac{3}{2}}}{9}\bigg[\left( \alpha+{\frac{15}{14}} \right)  \left( \beta\, \left( 1+z
 \right)  \right) ^{2\,\alpha}+ \left( \frac{3}{7}\,\alpha+{\frac{9}{14}}
 \right)  \left( \beta\, \left( 1+z \right)  \right) ^{4\,\alpha}\\&+&
 \left( \frac{\alpha}{7}-\frac{3}{14} \right)  \left( \beta\, \left( 1+z \right)
 \right) ^{6\,\alpha}+ \left( \frac{\alpha}{7}-\frac{3}{14} \right)  \left( \beta\,
 \left( 1+z \right)  \right) ^{8\,\alpha}+\frac{4}{7}\,\alpha+\frac{3}{7}\bigg]{{H_0}}^{2}{{\rm e}^{2\,{\frac { \left( 1+{\beta}^{2\,\alpha}\right) ^{\frac{3}{2}}{\beta}^{-2\,\alpha}}{\alpha\,{H_0}\,\sqrt {1+
 \left( \beta\, \left( 1+z \right)  \right) ^{2\,\alpha}}}}}}\\&+&{\frac {28\, \left( 1+{\beta}^{2\,\alpha} \right) ^{\frac{3}{2}} \left( 1+z\right) ^{2\,\alpha}}{9}}\bigg[\left( \alpha+{\frac{45}{28}} \right)  \left( \beta\, \left( 1+z
 \right)  \right) ^{2\,\alpha}+ \left( \frac{3}{7}\,\alpha+{\frac{27}{28}}
 \right)  \left( \beta\, \left( 1+z \right)  \right) ^{4\,\alpha}\\&+&
 \left( \frac{\alpha}{7}-{\frac{9}{28}} \right)  \left( \beta\, \left( 1+z
 \right)  \right) ^{6\,\alpha}+ \left( \frac{\alpha}{7}-{\frac{9}{28}}
 \right)  \left( \beta\, \left( 1+z \right)  \right) ^{8\,\alpha}+\frac{4}{7}
\,\alpha+{\frac{9}{14}}\bigg]{H_0}\,{{\rm e}^{{\frac { \left( 1+{\beta}^{2\,\alpha} \right) ^{\frac{3}{2}}{\beta}^{-2\,\alpha}}{\alpha\,{H_0}\,\sqrt {1+ \left( \beta\,
 \left( 1+z \right)  \right) ^{2\,\alpha}}}}}}\\&+&\left(  \left( \beta\, \left( 1+z \right)  \right) ^{2\,\alpha}-2
 \right)  \left( 1+ \left( \beta\, \left( 1+z \right)  \right) ^{2\,
\alpha} \right) ^{\frac{3}{2}} \left( 1+z \right) ^{4\,\alpha} \left( {\beta}^
{2\,\alpha}+{\beta}^{4\,\alpha}+\frac{1}{3}\,{\beta}^{6\,\alpha}+\frac{1}{3} \right),
\end{eqnarray*}
\begin{eqnarray*}
\triangle_{4}&=&\left( 1+ \left( \beta\, \left( 1+z \right)  \right) ^{2\,\alpha}
 \right) ^{\frac{5}{2}}\bigg[\left( 1+ \left( \beta\, \left( 1+z \right)  \right) ^{2\,\alpha}
 \right) ^{\frac{3}{2}}{H_0}\,{{\rm e}^{{\frac { \left( 1+{\beta}^{2\,
\alpha} \right) ^{\frac{3}{2}}{\beta}^{-2\,\alpha}}{\alpha\,{H_0}\,\sqrt {1
+ \left( \beta\, \left( 1+z \right)  \right) ^{2\,\alpha}}}}}}-\frac{3}{2}\,
 \left( 1+{\beta}^{2\,\alpha} \right) ^{\frac{3}{2}}\\&.& \left( 1+z \right) ^{2\,
\alpha}\bigg]\bigg[\left(  \left( \alpha-\frac{3}{2} \right)  \left( \beta\, \left( 1+z \right)
 \right) ^{2\,\alpha}-2\,\alpha-\frac{3}{2} \right) \sqrt {1+ \left( \beta\,
 \left( 1+z \right)  \right) ^{2\,\alpha}}{H_0}\,\\&.&{{\rm e}^{{\frac {
 \left( 1+{\beta}^{2\,\alpha} \right) ^{\frac{3}{2}}{\beta}^{-2\,\alpha}}{
\alpha\,{H_0}\,\sqrt {1+ \left( \beta\, \left( 1+z \right)
 \right) ^{2\,\alpha}}}}}}+\frac{3}{2}\, \left( 1+{\beta}^{2\,\alpha} \right)
^{\frac{3}{2}} \left( 1+z \right) ^{2\,\alpha}\bigg].
\end{eqnarray*}

\begin{figure*}[htp]
   \begin{minipage}{0.49\textwidth}
     \centering
   \includegraphics[scale=0.7]{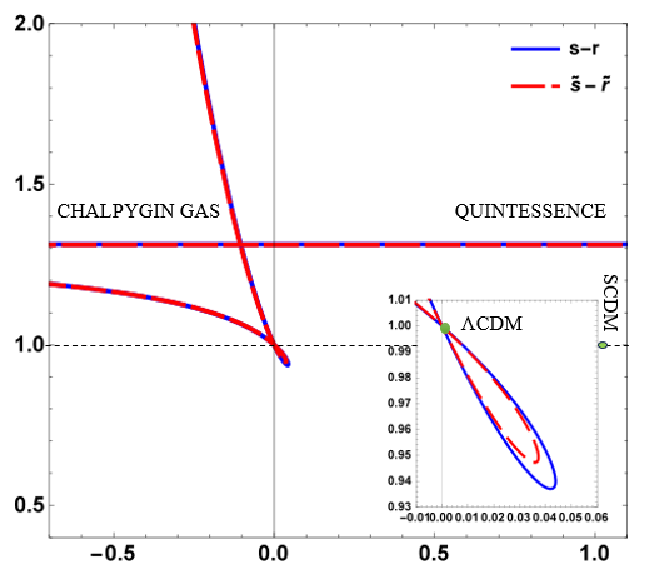}
\caption{Model 1: The $\tilde{\mathfrak{s}}-\tilde{r}$ plots show the different trajectories of the models in Riemannian and Finslerian modes.}\label{fig_17}
   \end{minipage}\hfill
   \begin{minipage}{0.49\textwidth}
     \centering
    \includegraphics[scale=0.7]{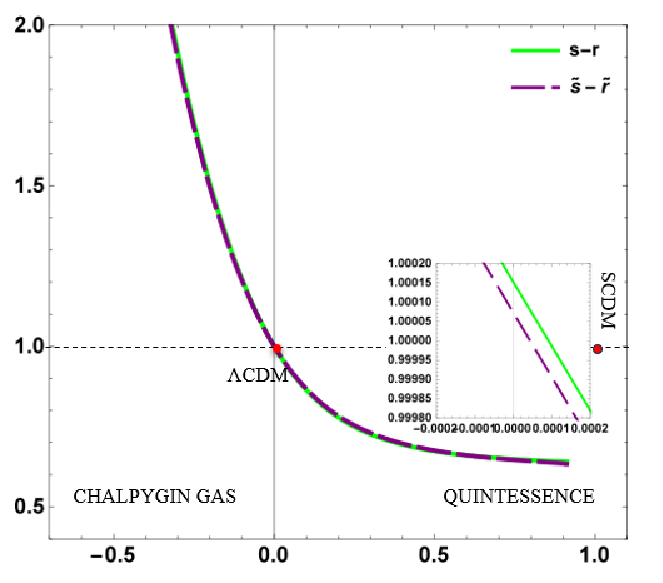}
\caption{Model 2: The $\tilde{\mathfrak{s}}-\tilde{r}$ plots show the different trajectories of the models in Riemannian and Finslerian modes.}\label{fig_18}
   \end{minipage}
\end{figure*}
\begin{figure*}[htp]
   \begin{minipage}{0.49\textwidth}
     \centering
   \includegraphics[scale=0.7]{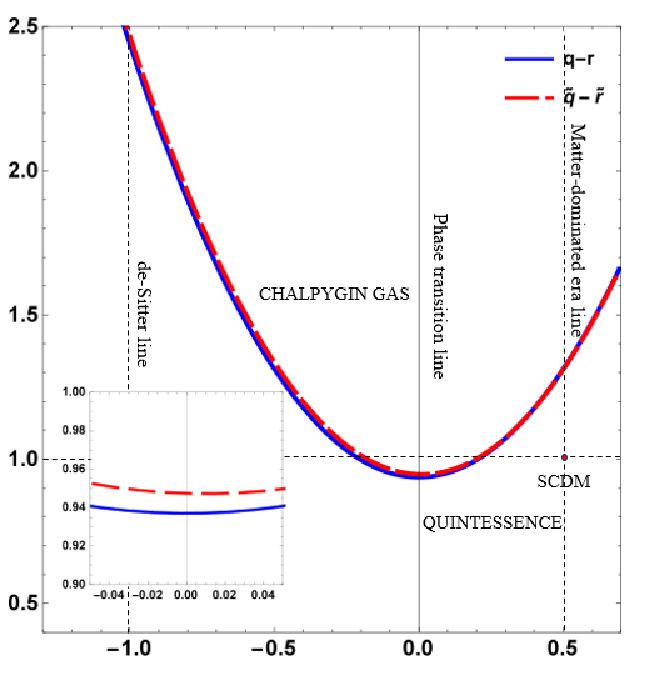}
\caption{Model 1: The $\tilde{\mathfrak{q}}-\tilde{r}$ plots show the different trajectories of the models in Riemannian and Finslerian modes.}\label{fig_19}
   \end{minipage}\hfill
   \begin{minipage}{0.49\textwidth}
     \centering
    \includegraphics[scale=0.7]{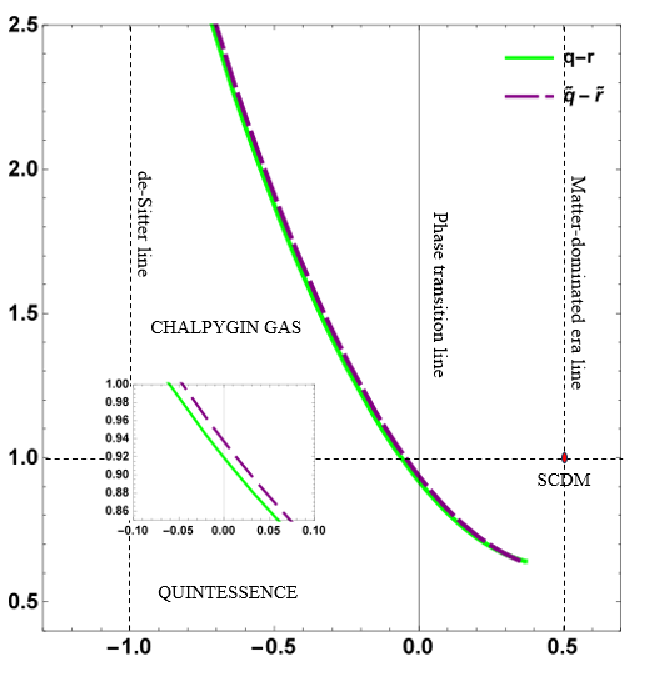}
\caption{Model 2: The $\tilde{\mathfrak{q}}-\tilde{r}$ plots show the different trajectories of the models in Riemannian and Finslerian modes.}\label{fig_20}
   \end{minipage}
\end{figure*}
\section{$Om(z)$ Diagnostic}
$Om$ diagnostic is another instrument defined in \cite{Sahni1, Zunckel, Shahalam, Agarwal} that uses the Hubble parameter and provides a null test of the $\Lambda$CDM model. Similar to the statefinder diagnostic, the $\tilde{O}m$ diagnostic is also a productive method to distinguish different DE models from the $\Lambda$CDM model according to the slope variation of $\tilde{O}m(z)$. The positive slope of the diagnostic indicates a quintessence nature ($\omega>-1$), the Negative slope of the diagnostic indicates a phantom nature ($\omega <-1$), and the constant slope with respect to redshift reveals the nature of dark energy coincides with the cosmological constant ($\omega =-1$).
The $\tilde{O}m(z)$ for a flat universe is defined as follows:

\begin{equation*}
\tilde{O}m(z)=\frac{\left(\frac{\tilde{H}(z)}{H_0}\right)^2-1}{(1+z)^3-1},
\end{equation*}
\textbf{Model 1}\\
\begin{equation*}
\tilde{O}m(z)=\frac{\left(\frac{\left(1+\left[\beta(1+z)\right]^\alpha\right)^4}
{\left(1+\beta^\alpha\right)^4\left(1+z\right)^{2\alpha}}-\frac{\left(1+\left[\beta(1+z)\right]^\alpha\right)^2}
{H_0\left(1+\beta^\alpha\right)^2\left(1+z\right)^{2\alpha}}\rm{e}^{-\frac{\left(1+\beta^\alpha\right)^2}{\alpha H_0\beta^\alpha\left(1+\left[\beta(1+z)\right]^\alpha\right)}}\right)-1}{(1+z)^3-1},
\end{equation*}
\textbf{Model 2}\\
\begin{equation*}
\tilde{O}m(z)=\frac{\left(\frac{\left(1+\left[\beta(1+z)\right]^{2\alpha}\right)^{3}}
{\left(1+\beta^{2\alpha}\right)^3\left(1+z\right)^{4\alpha}}
-\frac{\left(1+\left[\beta(1+z)\right]^{2\alpha}\right)^{\frac{3}{2}}}
{H_0\left(1+\beta^{2\alpha}\right)^\frac{3}{2}\left(1+z\right)^{2\alpha}}\rm{e}^{-\frac{\left(1+\beta^{2\alpha}\right)^\frac{3}{2}}{\alpha H_0\beta^{2\alpha}\left(1+\left[\beta(1+z)\right]^{2\alpha}\right)^\frac{1}{2}}}\right)-1}{(1+z)^3-1}.
\end{equation*}

\begin{figure*}[htp]
   \begin{minipage}{0.49\textwidth}
     \centering
   \includegraphics[scale=0.9]{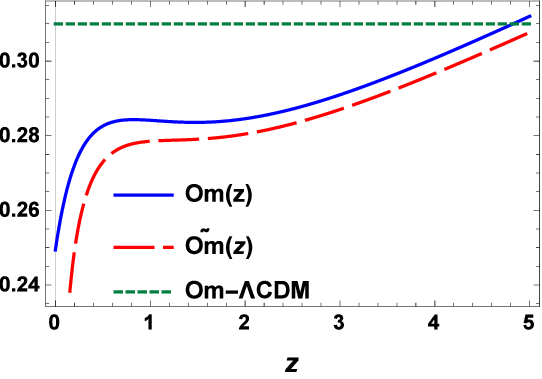}
\caption{$Om(z)$ vs. $z$ in Riemannian and Finslerian modes.}\label{fig11}
   \end{minipage}\hfill
   \begin{minipage}{0.49\textwidth}
     \centering
    \includegraphics[scale=0.9]{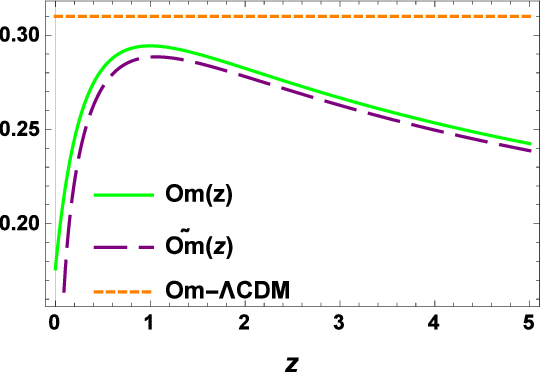}
\caption{$Om(z)$ vs. $z$ in Riemannian and Finslerian modes.}\label{fig12}
   \end{minipage}
\end{figure*}
\section{Physical Parameters}
Here, we consider a two-fluid model, cold dark matter and dark energy, since the radiation contribution at present is negligible. The matter pressure is $P_\mathfrak{m} = 0$ for cold dark matter, the equation of state is
$P_{\mathfrak{de}} =\omega_{\mathfrak{de}}\rho_{\mathfrak{de}}$ for dark energy, and $M_{pl}=(8\pi G)^{-\frac{1}{2}}$.
By utilizing Eqs (\ref{Eq22}) and (\ref{Eq23}), we derive expressions for critical physical parameters. These expressions allow us to interpret the physical behavior of matter and dark energy densities and pressures. Graphical representations further elucidate these behaviors for both Model 1 and Model 2, providing insights into their respective dynamics within the cosmological framework.\\\\
\textbf{Model 1:}\\
\begin{eqnarray}\label{Eq32}
\frac{\tilde{P}_{\mathfrak{de}}}{M_{pl}^2H_{0}^{2}}&=&\frac{(-3+2\alpha)\left(1+\beta^\alpha(1+z)^{\alpha}\right)^4
-4\alpha\left(1+\beta^\alpha(1+z)^{\alpha}\right)^3}{\left(1+\beta^\alpha\right)^4(1+z)^{2\alpha}}+5\frac{\left(1+\beta^\alpha(1+z)^{\alpha}\right)^2}
{H_0\left(1+\beta^\alpha\right)^2(1+z)^{\alpha}}\nonumber\\&.&\rm{e}^{\frac{-\left(1+\beta^\alpha\right)^2}{\alpha H_0\beta^\alpha\left(1+\left[\beta(1+z)\right]^\alpha\right)}},
\end{eqnarray}
\begingroup\makeatletter\def\f@size{10.5}\check@mathfonts
\begin{eqnarray}\label{Eq33}
\frac{\tilde{\rho}_{\mathfrak{de}}}{M_{pl}^2H_{0}^{2}}&=&\frac{1}{\omega_{\mathfrak{de}}}\bigg\{\frac{(-3+2\alpha)\left(1+\beta^\alpha(1+z)^{\alpha}\right)^4
-4\alpha\left(1+\beta^\alpha(1+z)^{\alpha}\right)^3}{\left(1+\beta^\alpha\right)^4(1+z)^{2\alpha}}+5\frac{\left(1+\beta^\alpha(1+z)^{\alpha}\right)^2}
{H_0\left(1+\beta^\alpha\right)^2(1+z)^{\alpha}}\nonumber\\&.&\rm{e}^{\frac{-\left(1+\beta^\alpha\right)^2}{\alpha H_0\beta^\alpha\left(1+\left[\beta(1+z)\right]^\alpha\right)}}\bigg\},
\end{eqnarray}
\endgroup
\begingroup\makeatletter\def\f@size{10.5}\check@mathfonts
\begin{eqnarray}\label{Eq34}
\frac{\tilde{\rho}_{\mathfrak{m}}}{M_{pl}^2H_{0}^{2}}&=&\left(3\frac{\left(1+\beta^\alpha(1+z)^{\alpha}\right)^4}
{\left(1+\beta^\alpha\right)^4(1+z)^{2\alpha}}-\frac{1}{\omega_{\mathfrak{de}}}\frac{(-3+2\alpha)\left(1+\beta^\alpha(1+z)^{\alpha}\right)^4
-4\alpha\left(1+\beta^\alpha(1+z)^{\alpha}\right)^3}{\left(1+\beta^\alpha\right)^4(1+z)^{2\alpha}}\right)\nonumber\\&-&(3+\frac{5}{\omega_{\mathfrak{de}}})\frac{\left(1+\beta^\alpha(1+z)^{\alpha}\right)^2}
{H_0\left(1+\beta^\alpha\right)^2(1+z)^{\alpha}}\rm{e}^{\frac{-\left(1+\beta^\alpha\right)^2}{\alpha H_0\beta^\alpha\left(1+\left[\beta(1+z)\right]^\alpha\right)}}.
\end{eqnarray}
\endgroup
\textbf{Model 2:}\\
\begingroup\makeatletter\def\f@size{11}\check@mathfonts
\begin{eqnarray}\label{Eq35}
\frac{\tilde{P}_{\mathfrak{de}}}{M_{pl}^2H_{0}^{2}}&=&\frac{(-3+2\alpha)\left(1+\beta^{2\alpha}(1+z)^{2\alpha}\right)^3
-6\alpha\left(1+\beta^{2\alpha}(1+z)^{2\alpha}\right)^2}{\left(1+\beta^{2\alpha}\right)^3(1+z)^{4\alpha}}+5\frac{\left(1+\beta^{2\alpha}(1+z)^{2\alpha}\right)^\frac{3}{2}}{H_0\left(1+\beta^{2\alpha}\right)^\frac{3}{2}(1+z)^{2\alpha}}
\nonumber\\&.&\rm{e}^{-\frac{\left(1+\beta^{2\alpha}\right)^\frac{3}{2}}{\alpha H_0\beta^{2\alpha}\left(1+\left[\beta(1+z)\right]^{2\alpha}\right)^\frac{1}{2}}},
\end{eqnarray}
\endgroup
\begingroup\makeatletter\def\f@size{10.3}\check@mathfonts
\begin{eqnarray}\label{Eq36}
\frac{\tilde{\rho}_{\mathfrak{de}}}{M_{pl}^2H_{0}^{2}}&=&\frac{1}{\omega_{\mathfrak{de}}}\bigg\{\frac{(-3+2\alpha)\left(1+\beta^{2\alpha}(1+z)^{2\alpha}\right)^3
-6\alpha\left(1+\beta^{2\alpha}(1+z)^{2\alpha}\right)^2}{\left(1+\beta^{2\alpha}\right)^3(1+z)^{4\alpha}}+5\frac{\left(1+\beta^{2\alpha}(1+z)^{2\alpha}\right)^\frac{3}{2}}{H_0\left(1+\beta^{2\alpha}\right)^\frac{3}{2}(1+z)^{2\alpha}}
\nonumber\\&.&\rm{e}^{-\frac{\left(1+\beta^{2\alpha}\right)^\frac{3}{2}}{\alpha H_0\beta^{2\alpha}\left(1+\left[\beta(1+z)\right]^{2\alpha}\right)^\frac{1}{2}}}\bigg\},
\end{eqnarray}
\endgroup
\begingroup\makeatletter\def\f@size{10.5}\check@mathfonts
\begin{eqnarray}\label{Eq37}
\frac{\tilde{\rho}_{\mathfrak{m}}}{M_{pl}^2H_{0}^{2}}&=&\left(3\frac{\left(1+\beta^{2\alpha}(1+z)^{2\alpha}\right)^3}{\left(1+\beta^{2\alpha}\right)^3(1+z)^{4\alpha}}-
\frac{1}{\omega_{}\mathfrak{de}}\frac{(-3+2\alpha)\left(1+\beta^{2\alpha}(1+z)^{2\alpha}\right)^3
-6\alpha\left(1+\beta^{2\alpha}(1+z)^{2\alpha}\right)^2}{\left(1+\beta^{2\alpha}\right)^3(1+z)^{4\alpha}}\right)\nonumber\\&-&(3+\frac{5}{\omega_{\mathfrak{de}}})\frac{\left(1+\beta^{2\alpha}(1+z)^{2\alpha}\right)^\frac{3}{2}}{H_0\left(1+\beta^{2\alpha}\right)^\frac{3}{2}(1+z)^{2\alpha}}
{e}^{-\frac{\left(1+\beta^{2\alpha}\right)^\frac{3}{2}}{\alpha H_0\beta^{2\alpha}\left(1+\left[\beta(1+z)\right]^{2\alpha}\right)^\frac{1}{2}}}.
\end{eqnarray}
\endgroup
When the candidate of dark energy is considered as cosmological constant, we have $\rho_{\mathfrak{de}}=\rho_{\Lambda}$ and $\omega_{\mathfrak{de}}=-1$, then the evolution of the physical parameters can be illustrated as shown in the following figures:
\begin{figure*}[htp]
   \begin{minipage}{0.49\textwidth}
     \centering
   \includegraphics[scale=0.55]{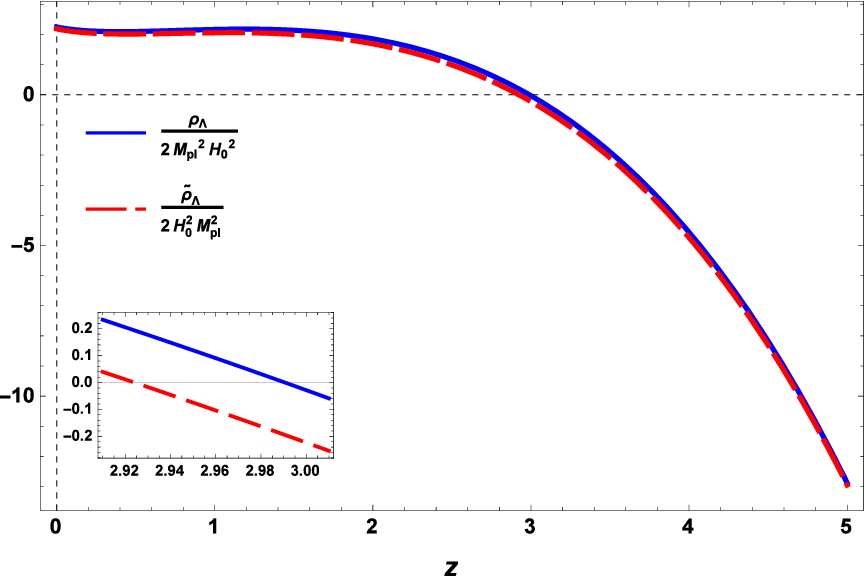}
\caption{Model 1: The evolution of the energy densities of the cosmological constant ($\rho_{\mathfrak{de}}=\rho_{\Lambda}$ and $\omega_{\mathfrak{de}}=-1$) in Riemannian and Finslerian modes.}\label{fig-PH1}
   \end{minipage}\hfill
   \begin{minipage}{0.49\textwidth}
     \centering
    \includegraphics[scale=0.55]{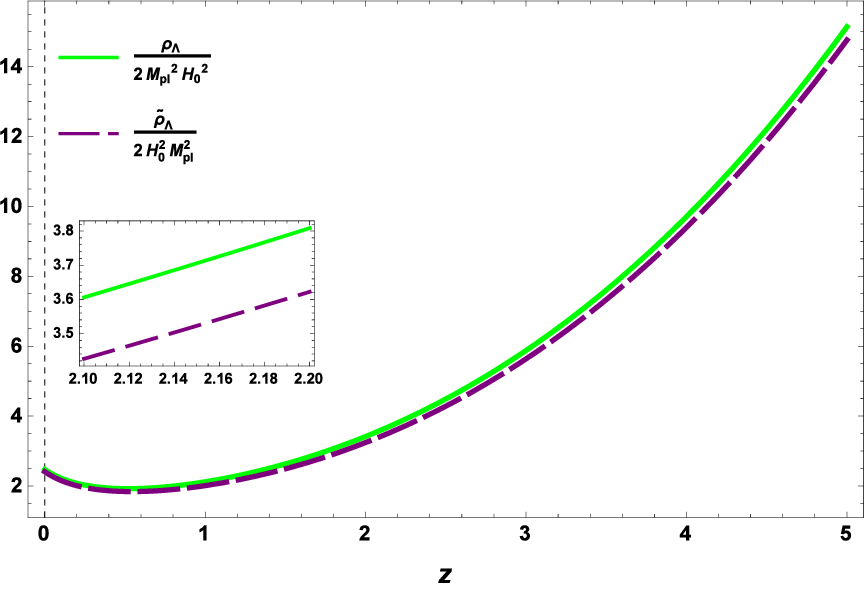}
\caption{Model 2: The evolution of the energy densities of the cosmological constant ($\rho_{\mathfrak{de}}=\rho_{\Lambda}$ and $\omega_{\mathfrak{de}}=-1$) in Riemannian and Finslerian modes.}\label{fig-PH2}
   \end{minipage}
\end{figure*}
\begin{figure*}[htp]
   \begin{minipage}{0.49\textwidth}
     \centering
   \includegraphics[scale=0.55]{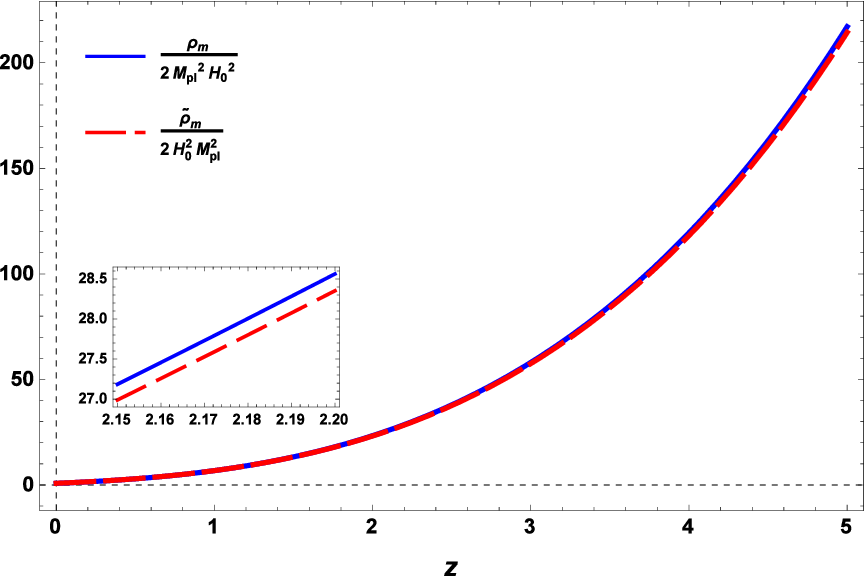}
\caption{Model 1: The evolution of the matter energy densities ($\rho_{m}$) in Riemannian and Finslerian modes.}\label{fig-PH3}
   \end{minipage}\hfill
   \begin{minipage}{0.49\textwidth}
     \centering
    \includegraphics[scale=0.55]{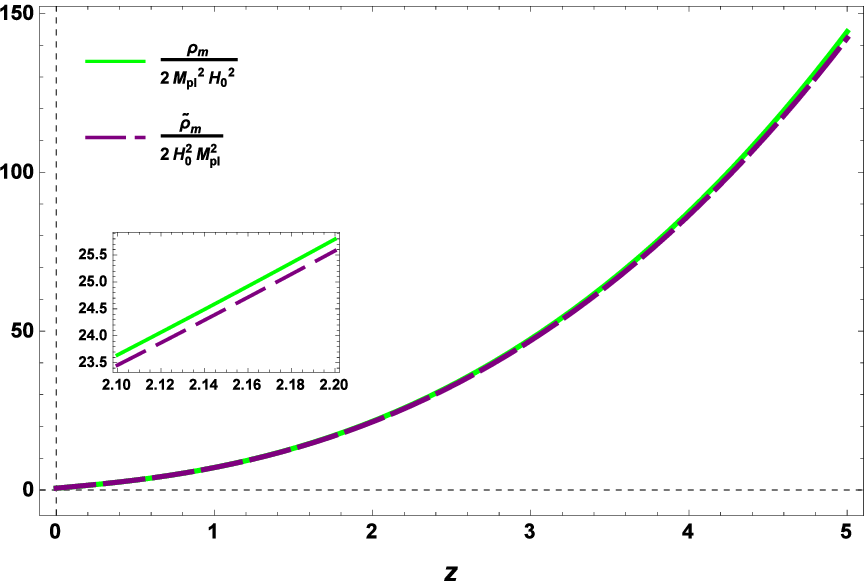}
\caption{Model 2: The evolution of the matter energy densities ($\rho_{m}$) in Riemannian and Finslerian modes.}\label{fig-PH4}
   \end{minipage}
\end{figure*}
\textbf{Scalar Field}
Since the equation of state for the cosmological constant is non-dynamical, the observations favor the dynamical equation of state. This feature makes another suitable candidate for depicting dark energy, such as the general scalar field. For a general scalar field $\phi$, the action can be denoted as
\begin{equation*}
S=\frac{1}{16\pi}\int d^4x\sqrt{-g}\left\{\frac{M_{pl}^{2}}{2}R-\frac{1}{2}\partial_{\nu}\partial^{\nu}\phi-V(\phi)+{}^{Matter}\mathcal{L}\right\}.
\end{equation*}
In 2018, S. I. Vacaru \cite{SVacaru} derived respective energy-momentum tensors for Lagrange-Finsler. Due to this
\begin{equation*}
S={}^{g}S+{}^{\phi}S+{}^{Matter}S, ~~~ T_{\alpha\beta}={}^{g}T_{\alpha\beta}+{}^{\phi}T_{\alpha\beta}.
\end{equation*}
In GR, the Lagrange density for gravitational fields is postulated in the form
\begin{equation*}
{}^{g}\mathcal{L}(g_{ij},\nabla)=\frac{M_{pl}^{2}}{2}R,
\end{equation*}
where $R$ is the Ricci scalar and the Planck mass $M_{pl}$ is determined by the Newton constant $G$. $R_{ij}=\frac{8\pi G}{c^4}\left({}^{g}T_{ij}-\frac{1}{2}g_{ij}{}^{g}T\right)$. The Lagrange density for scalar fields is defined in the following form
\begin{equation*}
{}^{\phi}\mathcal{L}=\frac{1}{2}\dot{\phi}^2-V(\phi),
\end{equation*}
where ${}^{\phi}T_{\alpha\beta}=\frac{-2}{\sqrt{|g_{\alpha\beta}|}}\frac{\delta(\sqrt{|g_{\mu\nu}|}{}^{\phi}\mathcal{L})}{\delta g^{\alpha\beta}}$.
Therefore, similar to the Riemannian state, we have
\begin{equation*}
\tilde{\rho}_{\phi}={}^{\phi}T_{00}=\frac{1}{2}\dot{\phi}^2+V(\phi).
\end{equation*}
For a two-component universe, scalar field and cold dark matter with minimal interaction between them (i.e., $\rho_{\mathfrak{tot}}=\rho_{\phi}+\rho_{\mathfrak{m}},
\rho_{\mathfrak{m}}= ca^{-3}= c(1 + z)^3$, $c$ is a constant of integration), then the solutions obtained from Eqs. (\ref{Eq22}) and (\ref{Eq23}) are\\\\
\textbf{Model 1:}\\
\begin{eqnarray}\label{Eq38}
\frac{\tilde{\rho}_{\phi}}{M_{pl}^2H_{0}^{2}}&=&\frac{c(1+z)^3}{M_{pl}^2H_{0}^{2}}-\left(\frac{2\alpha\left(1+\beta^\alpha(1+z)^{\alpha}\right)^4
-4\alpha\left(1+\beta^\alpha(1+z)^{\alpha}\right)^3}{\left(1+\beta^\alpha\right)^4(1+z)^{2\alpha}}\right)\nonumber\\&-&2\frac{\left(1+\beta^\alpha(1+z)^{\alpha}\right)^2}
{H_0\left(1+\beta^\alpha\right)^2(1+z)^{\alpha}}\rm{e}^{\frac{-\left(1+\beta^\alpha\right)^2}{\alpha H_0\beta^\alpha\left(1+\left[\beta(1+z)\right]^\alpha\right)}},
\end{eqnarray}
\begin{eqnarray}\label{Eq39}
\frac{\tilde{V}(\phi)}{M_{pl}^2H_{0}^{2}}&=&\frac{c(1+z)^3}{2M_{pl}^2H_{0}^{2}}-\left(\frac{(\frac{-3}{2}+2\alpha)\left(1+\beta^\alpha(1+z)^{\alpha}\right)^4
-4\alpha\left(1+\beta^\alpha(1+z)^{\alpha}\right)^3}{\left(1+\beta^\alpha\right)^4(1+z)^{2\alpha}}\right)\nonumber\\&-&\frac{7}{2}\frac{\left(1+\beta^\alpha(1+z)^{\alpha}\right)^2}
{H_0\left(1+\beta^\alpha\right)^2(1+z)^{\alpha}}\rm{e}^{\frac{-\left(1+\beta^\alpha\right)^2}{\alpha H_0\beta^\alpha\left(1+\left[\beta(1+z)\right]^\alpha\right)}}.
\end{eqnarray}
\textbf{Model 2:}\\
\begin{eqnarray}\label{Eq40}
\frac{\tilde{\rho}_{\phi}}{M_{pl}^2H_{0}^{2}}&=&\frac{c(1+z)^3}{M_{pl}^2H_{0}^{2}}-\left(\frac{2\alpha\left(1+\beta^{2\alpha}(1+z)^{2\alpha}\right)^3
-6\alpha\left(1+\beta^{2\alpha}(1+z)^{2\alpha}\right)^2}{\left(1+\beta^{2\alpha}\right)^3(1+z)^{4\alpha}}\right)\nonumber\\&-&2\frac{\left(1+\beta^{2\alpha}(1+z)^{2\alpha}\right)^\frac{3}{2}}{H_0\left(1+\beta^{2\alpha}\right)^\frac{3}{2}(1+z)^{2\alpha}}
\rm{e}^{-\frac{\left(1+\beta^{2\alpha}\right)^\frac{3}{2}}{\alpha H_0\beta^{2\alpha}\left(1+\left[\beta(1+z)\right]^{2\alpha}\right)^\frac{1}{2}}},
\end{eqnarray}
\begin{eqnarray}\label{Eq41}
\frac{\tilde{V}(\phi)}{M_{pl}^2H_{0}^{2}}&=&\frac{c(1+z)^3}{2M_{pl}^2H_{0}^{2}}-\left(\frac{(\frac{-3}{2}+2\alpha)\left(1+\beta^{2\alpha}(1+z)^{2\alpha}\right)^3
-6\alpha\left(1+\beta^{2\alpha}(1+z)^{2\alpha}\right)^2}{\left(1+\beta^{2\alpha}\right)^3(1+z)^{4\alpha}}\right)\nonumber\\&-&\frac{7}{2}\frac{\left(1+\beta^{2\alpha}(1+z)^{2\alpha}\right)^\frac{3}{2}}{H_0\left(1+\beta^{2\alpha}\right)^\frac{3}{2}(1+z)^{2\alpha}}
\rm{e}^{-\frac{\left(1+\beta^{2\alpha}\right)^\frac{3}{2}}{\alpha H_0\beta^{2\alpha}\left(1+\left[\beta(1+z)\right]^{2\alpha}\right)^\frac{1}{2}}}.
\end{eqnarray}
\begin{figure*}[htp]
   \begin{minipage}{0.49\textwidth}
     \centering
   \includegraphics[scale=0.55]{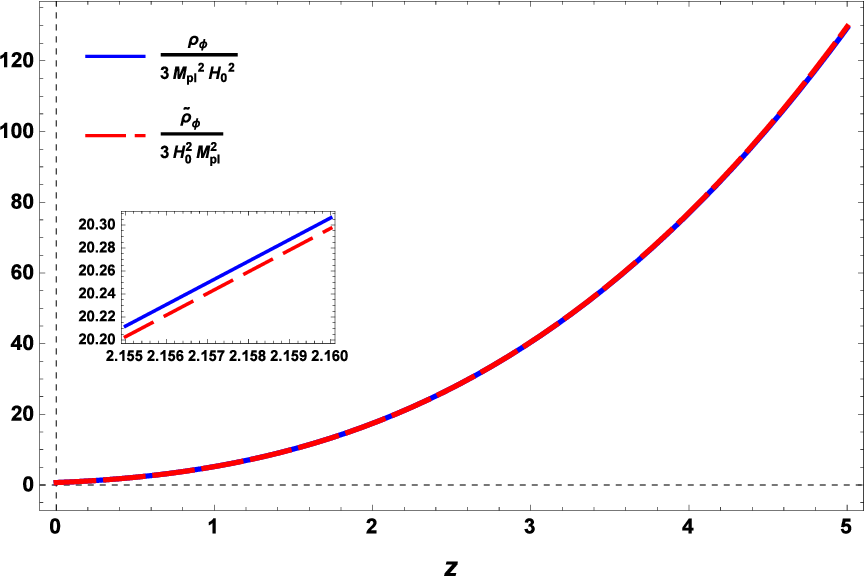}
\caption{Model 1:  The evolution of the scalar field energy density ($\rho_{\phi}$) in Riemannian and Finslerian modes.}\label{fig-PH5}
   \end{minipage}\hfill
   \begin{minipage}{0.49\textwidth}
     \centering
    \includegraphics[scale=0.55]{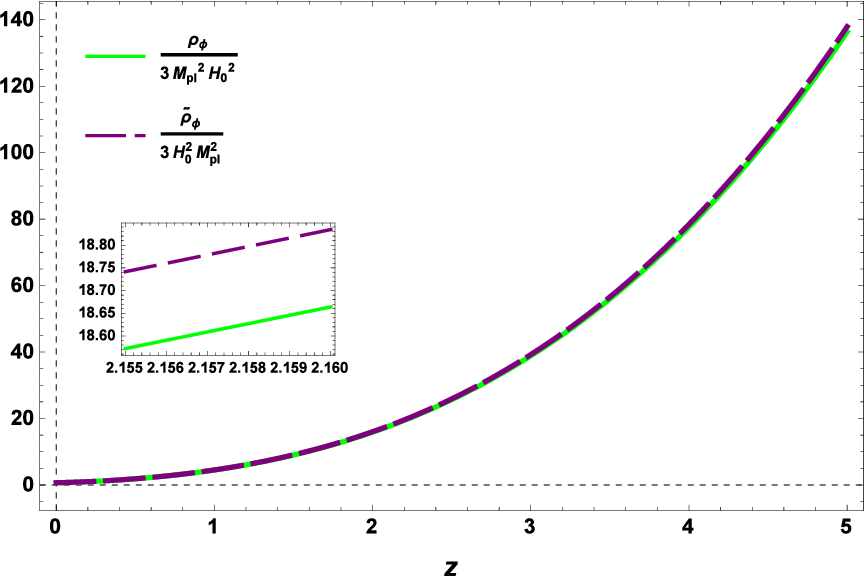}
\caption{Model 2: The evolution of the scalar field energy density ($\rho_{\phi}$) in Riemannian and Finslerian modes.}\label{fig-PH6}
   \end{minipage}
\end{figure*}
\begin{figure*}[htp]
   \begin{minipage}{0.49\textwidth}
     \centering
   \includegraphics[scale=0.55]{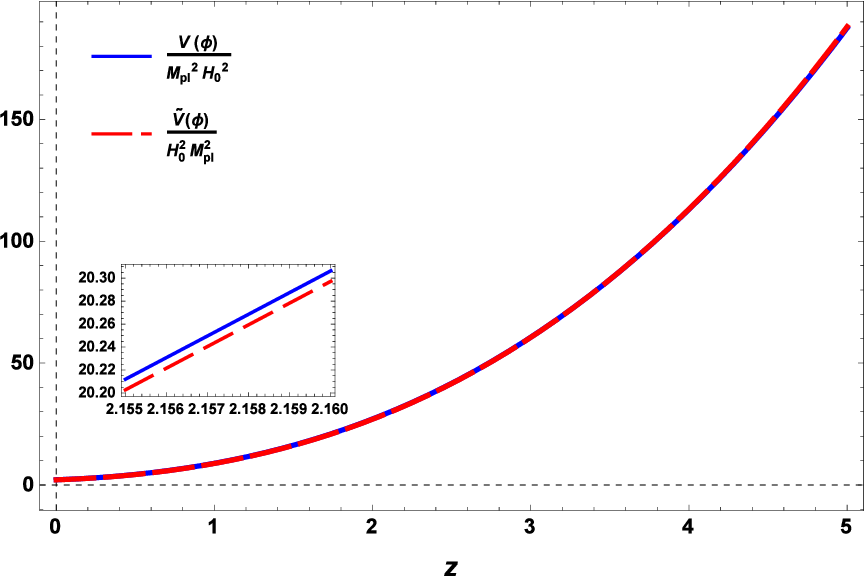}
\caption{Model 1: The evolution of the scalar field potential ($V(\phi)$) in Riemannian and Finslerian modes.}\label{fig-PH7}
   \end{minipage}\hfill
   \begin{minipage}{0.49\textwidth}
     \centering
    \includegraphics[scale=0.55]{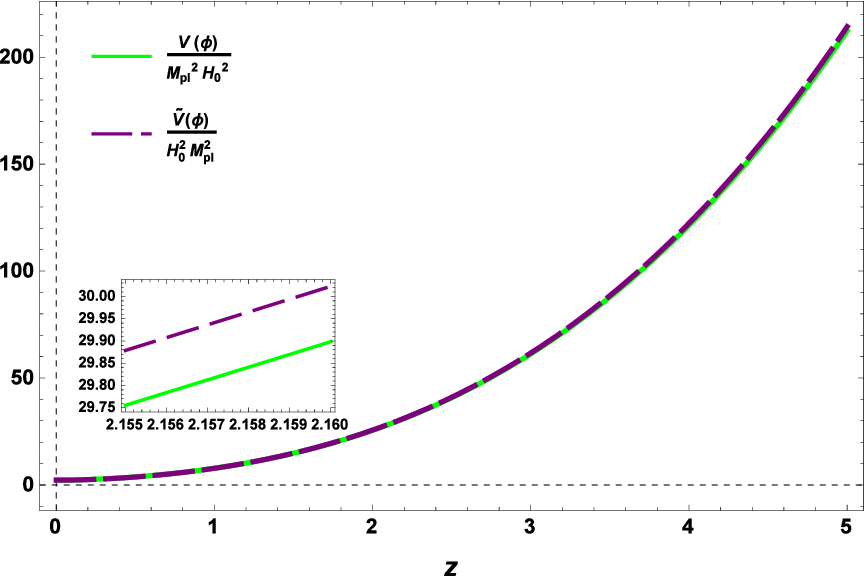}
\caption{Model 2: The evolution of the scalar field potential ($V(\phi)$) in Riemannian and Finslerian modes.}\label{fig-PH8}
   \end{minipage}
\end{figure*}
As we know, the total density parameters is written as follows,
\begin{equation}\label{Eq42}
1=\tilde{\Omega}=\tilde{\Omega}_{\phi}+\tilde{\Omega}_{\mathfrak{m}}
\end{equation}
where $\tilde{\Omega}_{\phi}=\frac{\tilde{\rho}_{\phi}}{3M_{pl}^{2}\tilde{H}^2}$ and $\tilde{\Omega}_{\mathfrak{m}}=\frac{\tilde{\rho}_{\mathfrak{m}}}{3M_{pl}^{2}\tilde{H}^2}$ are density parameter for the scalar field and density parameter for the matter, respectively. They can be computed for two models as,\\\\
\textbf{Model 1:}\\
\begin{equation}\label{Eq43}
\tilde{\Omega}_{\phi}=\frac{\frac{c(1+z)^3}{M_{pl}^2H_{0}^{2}}-\left(\frac{2\alpha\left(1+\beta^\alpha(1+z)^{\alpha}\right)^4
-4\alpha\left(1+\beta^\alpha(1+z)^{\alpha}\right)^3}{\left(1+\beta^\alpha\right)^4(1+z)^{2\alpha}}\right)-2\frac{\left(1+\beta^\alpha(1+z)^{\alpha}\right)^2}
{H_0\left(1+\beta^\alpha\right)^2(1+z)^{\alpha}}\rm{e}^{\frac{-\left(1+\beta^\alpha\right)^2}{\alpha H_0\beta^\alpha\left(1+\left[\beta(1+z)\right]^\alpha\right)}}
}{3\left(\frac{\left(1+\left[\beta(1+z)\right]^\alpha\right)^4}
{\left(1+\beta^\alpha\right)^4\left(1+z\right)^{2\alpha}}-\frac{\left(1+\left[\beta(1+z)\right]^\alpha\right)^2}
{H_0\left(1+\beta^\alpha\right)^2\left(1+z\right)^{2\alpha}}\rm{e}^{-\frac{\left(1+\beta^\alpha\right)^2}{\alpha H_0\beta^\alpha\left(1+\left[\beta(1+z)\right]^\alpha\right)}}\right)},
\end{equation}
from Eqs. (\ref{Eq42}) and (\ref{Eq43}), we obtain
\begin{equation}\label{Eq44}
\frac{c}{M_{pl}^{2}H_{0}^2}=3(1-\Omega_{\mathfrak{m}_0})(1-\frac{1}{H_0}\rm{e}^{\frac{-(1+\beta^\alpha)}{\alpha H_0\beta^\alpha}})+(2\alpha-\frac{4\alpha}{1+\beta^\alpha})+\frac{2}{H_0}\rm{e}^{\frac{-(1+\beta^\alpha)}{\alpha H_0\beta^\alpha}}=\mathcal{A},
\end{equation}
the equations of state parameter ($\tilde{\omega}_{\phi}=\frac{\tilde{P}_{\phi}}{\tilde{\rho}_{\phi}}$) are given by
\begin{equation}\label{Eq45}
\tilde{\omega}_{\phi}=\frac{\frac{(-3+2\alpha)\left(1+\beta^\alpha(1+z)^{\alpha}\right)^4
-4\alpha\left(1+\beta^\alpha(1+z)^{\alpha}\right)^3}{\left(1+\beta^\alpha\right)^4(1+z)^{2\alpha}}+5\frac{\left(1+\beta^\alpha(1+z)^{\alpha}\right)^2}
{H_0\left(1+\beta^\alpha\right)^2(1+z)^{\alpha}}\rm{e}^{\frac{-\left(1+\beta^\alpha\right)^2}{\alpha H_0\beta^\alpha\left(1+\left[\beta(1+z)\right]^\alpha\right)}}}{\mathcal{A}(1+z)^3-\left(\frac{2\alpha\left(1+\beta^\alpha(1+z)^{\alpha}\right)^4
-4\alpha\left(1+\beta^\alpha(1+z)^{\alpha}\right)^3}{\left(1+\beta^\alpha\right)^4(1+z)^{2\alpha}}\right)-2\frac{\left(1+\beta^\alpha(1+z)^{\alpha}\right)^2}
{H_0\left(1+\beta^\alpha\right)^2(1+z)^{\alpha}}\rm{e}^{\frac{-\left(1+\beta^\alpha\right)^2}{\alpha H_0\beta^\alpha\left(1+\left[\beta(1+z)\right]^\alpha\right)}}}.
\end{equation}
\textbf{Model 2:}\\
\begin{equation}\label{Eq46}
\tilde{\Omega}_{\phi}=\frac{\frac{c(1+z)^3}{M_{pl}^2H_{0}^{2}}-\left(\frac{2\alpha\left(1+\beta^{2\alpha}(1+z)^{2\alpha}\right)^3
-6\alpha\left(1+\beta^{2\alpha}(1+z)^{2\alpha}\right)^2}{\left(1+\beta^{2\alpha}\right)^3(1+z)^{4\alpha}}\right)-2\frac{\left(1+\beta^{2\alpha}(1+z)^{2\alpha}\right)^\frac{3}{2}}{H_0\left(1+\beta^{2\alpha}\right)^\frac{3}{2}(1+z)^{2\alpha}}
\rm{e}^{-\frac{\left(1+\beta^{2\alpha}\right)^\frac{3}{2}}{\alpha H_0\beta^{2\alpha}\left(1+\left[\beta(1+z)\right]^{2\alpha}\right)^\frac{1}{2}}}}{3\left(\frac{\left(1+\left[\beta(1+z)\right]^{2\alpha}\right)^{3}}
{\left(1+\beta^{2\alpha}\right)^3\left(1+z\right)^{4\alpha}}
-\frac{\left(1+\left[\beta(1+z)\right]^{2\alpha}\right)^{\frac{3}{2}}}
{H_0\left(1+\beta^{2\alpha}\right)^\frac{3}{2}\left(1+z\right)^{2\alpha}}\rm{e}^{-\frac{\left(1+\beta^{2\alpha}\right)^\frac{3}{2}}{\alpha H_0\beta^{2\alpha}\left(1+\left[\beta(1+z)\right]^{2\alpha}\right)^\frac{1}{2}}}\right)},
\end{equation}
from Eqs. (\ref{Eq42}) and (\ref{Eq46}), we obtain
\begin{equation}\label{Eq47}
\frac{c}{M_{pl}^{2}H_{0}^2}=3(1-\Omega_{\mathfrak{m}_0})(1-\frac{1}{H_0}\rm{e}^{\frac{-(1+\beta^{2\alpha})}{\alpha H_0\beta^{2\alpha}}})+(2\alpha-\frac{6\alpha}{1+\beta^{2\alpha}})+\frac{2}{H_0}\rm{e}^{\frac{-(1+\beta^{2\alpha})}{\alpha H_0\beta^{2\alpha}}}=\mathcal{A},
\end{equation}
 the equations of state parameter ($\tilde{\omega}_{\phi}=\frac{\tilde{P}_{\phi}}{\tilde{\rho}_{\phi}}$) are given by
\begin{equation}\label{Eq48}
\tilde{\omega}_{\phi}=\frac{\frac{(-3+2\alpha)\left(1+\beta^{2\alpha}(1+z)^{2\alpha}\right)^3
-6\alpha\left(1+\beta^{2\alpha}(1+z)^{2\alpha}\right)^2}{\left(1+\beta^{2\alpha}\right)^3(1+z)^{4\alpha}}+5\frac{\left(1+\beta^{2\alpha}(1+z)^{2\alpha}\right)^\frac{3}{2}}{H_0\left(1+\beta^{2\alpha}\right)^\frac{3}{2}(1+z)^{2\alpha}}
\rm{e}^{-\frac{\left(1+\beta^{2\alpha}\right)^\frac{3}{2}}{\alpha H_0\beta^{2\alpha}\left(1+\left[\beta(1+z)\right]^{2\alpha}\right)^\frac{1}{2}}}}{\mathcal{A}(1+z)^3-\left(\frac{2\alpha\left(1+\beta^{2\alpha}(1+z)^{2\alpha}\right)^3
-6\alpha\left(1+\beta^{2\alpha}(1+z)^{2\alpha}\right)^2}{\left(1+\beta^{2\alpha}\right)^3(1+z)^{4\alpha}}\right)-2\frac{\left(1+\beta^{2\alpha}(1+z)^{2\alpha}\right)^\frac{3}{2}}{H_0\left(1+\beta^{2\alpha}\right)^\frac{3}{2}(1+z)^{2\alpha}}
\rm{e}^{-\frac{\left(1+\beta^{2\alpha}\right)^\frac{3}{2}}{\alpha H_0\beta^{2\alpha}\left(1+\left[\beta(1+z)\right]^{2\alpha}\right)^\frac{1}{2}}}}.
\end{equation}
\begin{figure*}[htp]
   \begin{minipage}{0.49\textwidth}
     \centering
   \includegraphics[scale=0.55]{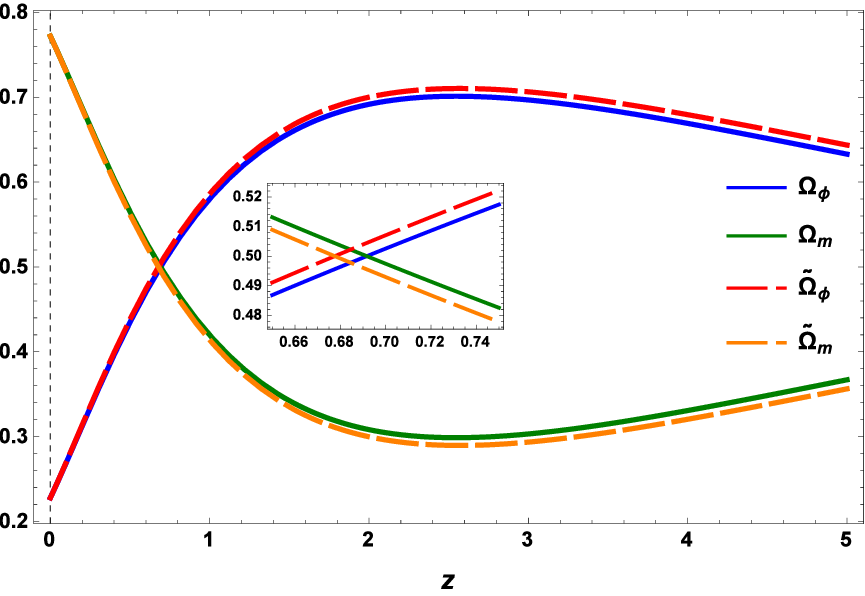}
\caption{Model 1: The evolution of the density parameters $\Omega_{\phi}$ and $\Omega_{\mathfrak{m}}$  in Riemannian and Finslerian modes.}\label{fig-PH9}
   \end{minipage}\hfill
   \begin{minipage}{0.49\textwidth}
     \centering
    \includegraphics[scale=0.55]{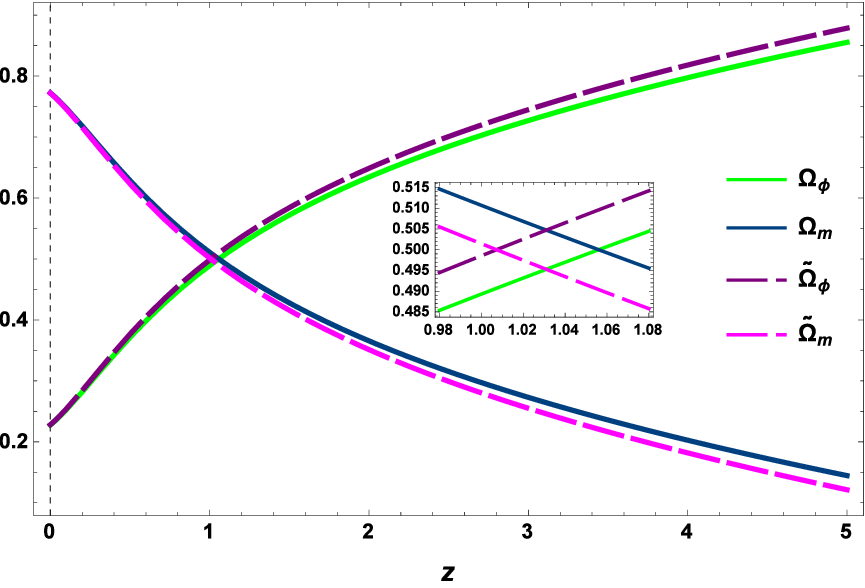}
\caption{Model 2: The evolution of the density parameters $\Omega_{\phi}$ and $\Omega_{\mathfrak{m}}$ in Riemannian and Finslerian modes.}\label{fig-PH10}
   \end{minipage}
\end{figure*}
\begin{figure*}[htp]
   \begin{minipage}{0.49\textwidth}
     \centering
   \includegraphics[scale=0.55]{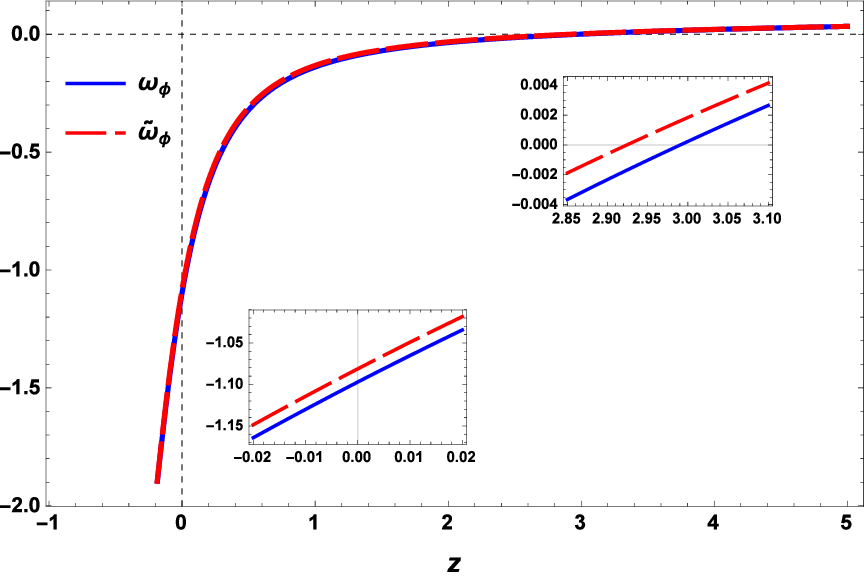}
\caption{Model 1: The evolution of equation of state parameter $\omega_\phi$ in Riemannian and Finslerian modes.}\label{fig-PH11}
   \end{minipage}\hfill
   \begin{minipage}{0.49\textwidth}
     \centering
    \includegraphics[scale=0.55]{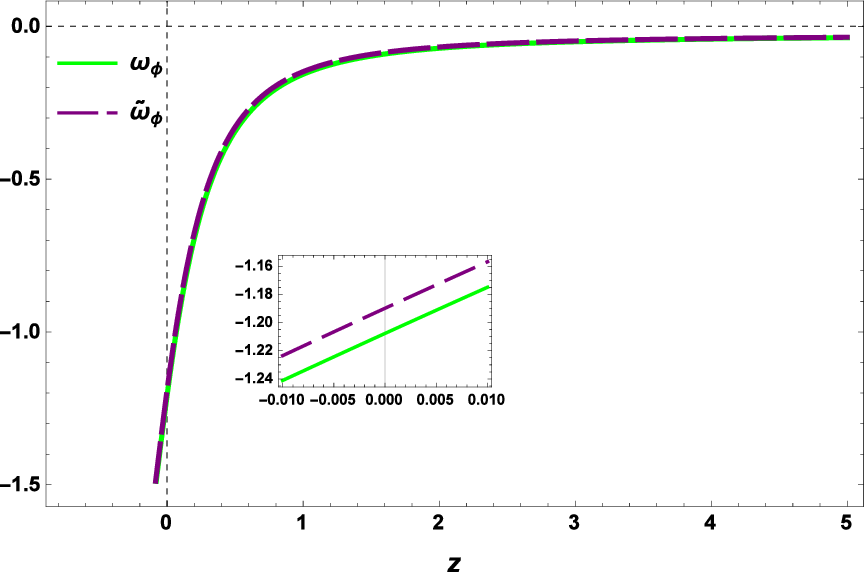}
\caption{Model 2: The evolution of equation of state parameter $\omega_\phi$ in Riemannian and Finslerian modes.}\label{fig-PH12}
   \end{minipage}
\end{figure*}
\section{Statistical Analysis}
When comparing the dark energy models in both Riemannian and Finslerian modes with the $\Lambda$CDM model, it's important to recognize that the $\Lambda$CDM model is encompassed within both dark energy models. This nesting property allows for the application of standard statistical tests. We compute these tests using the complete dataset. Consequently, one can define the reduced chi-square statistic as $\chi^2_{\text{red}} = \frac{\chi^2}{\text{DoF}}$, where \(\text{DoF}\) represents the degrees of freedom of the model, and \(\chi^2\) denotes the weighted sum of squared deviations. This calculation is performed under the constraint of the same number of runs. Certainly, here's an addition to the text that includes the P test \cite{118,119,120}: The Akaike information criterion (AIC) \cite{114,115} provides a means of comparing different models, considering a large dataset (in our case \(N = 273\)). It is defined as follows: $\text{AIC} = -2\ln(\mathcal{L}_{\text{max}}) + 2k.$ We utilize the Akaike information criterion (AIC) and the Bayesian information criterion (BIC) to evaluate our cosmological models. Specifically, the AIC is defined as:
\begin{equation}
\text{AIC} = -2\ln(\mathcal{L}_{\text{max}}) + 2k + \frac{2k(2k + 1)}{N_{\text{tot}} - k - 1},
\end{equation}
where \(\mathcal{L}_{\text{max}}\) represents the maximum likelihood of the considered data, \(N_{\text{tot}}\) is the total number of data points, and \(k\) denotes the number of parameters. For large \(N_{\text{tot}}\), the expression simplifies to:
\begin{equation}
\text{AIC} \approx -2\ln(\mathcal{L}_{\text{max}}) + 2k,
\end{equation}
which is the standard form of the AIC criterion. Conversely, the Bayesian information criterion is defined as:
\begin{equation}
\text{BIC} = -2\ln(\mathcal{L}_{\text{max}}) + k\ln N_{\text{tot}}.
\end{equation}
In addition to AIC and BIC, we can also consider the P-test, which evaluates the significance of the difference between the two models. This test quantifies whether the improvement in fit provided by a more complex model (compared to a simpler one) is statistically significant. It is commonly used in model selection to determine whether additional parameters introduced in a more intricate model are justified by improved fit. The P-test complements AIC and BIC in providing model performance and complexity with a comprehensive assessment.
\begin{table*}
\begin{center}
\begin{tabular}{|c|c|c|c|c|c|}
\hline
Model & \(\mathcal{L}_{\text{max}}\) & $\chi_{red}^{2}$ & $AIC$ & BIC & P-Value \\ \hline
$\Lambda$CDM Model & 254.53 & 0.9497 & 264.53 & 264.76 &  0.7127 \\ \hline
Model 1 Finslerian Geometry & 253.32 & 0.9452 & 263.32 & 263.55 & 0.7307 \\ \hline
Model 1 Riemannian Geometry & 253.60 & 0.9462 & 263.60 & 263.82 & 0.7267 \\ \hline
Model 2 Finslerian Geometry & 250.75 & 0.9356 & 260.68 & 260.90 & 0.7675 \\ \hline
Model 2 Riemannian Geometry & 250.68 & 0.9353 & 260.75 & 260.98 & 0.7686 \\ \hline
\end{tabular}%
\end{center}
\caption{Summary of the \(\mathcal{L}_{\text{max}}\), $\chi_{red}^{2}$, AIC, BIC and P-Value}
\label{tab_11}
\end{table*}
\section{Results and Discussion}
\paragraph{Deceleration parameter}
Figs. \ref{fig_9} and \ref{fig_10} illustrate the evolution of the deceleration parameter (\(q\)) in terms of redshift for both Model 1 and Model 2, in both Riemannian and Finslerian geometries, compared to the standard \(\Lambda\)CDM model. In Fig. \ref{fig_9}, one can observe that in the distant past (\(z \rightarrow \infty\)), Model 1 predicts values of approximately 0.6932 in Finslerian geometry and 0.6921 in Riemannian geometry. In contrast, the \(\Lambda\)CDM model predicts a value around 0.449. This behavior suggests that in the early Universe, both Model 1 variants exhibit higher deceleration compared to the \(\Lambda\)CDM model. As the redshift approaches zero (\(z \rightarrow 0\)), the present-day deceleration parameter (\(q_{0}\)) is approximately -0.6238 in Finslerian geometry and -0.6227 in Riemannian geometry. In contrast, the \(\Lambda\)CDM model predicts a value around -0.5506.
This indicates that, in the current epoch, both geometrical models (Finslerian and Riemannian) predict a slightly higher Universe acceleration compared to the \(\Lambda\)CDM model. The phase transition redshift (\(z_{tr}\)) is 0.6958 in Finslerian geometry, 0.7066 in Riemannian geometry, and 0.6824 in the \(\Lambda\)CDM model. The phase transition redshift marks the point at which the Universe transitions from decelerated to accelerated expansion. In Finslerian and Riemannian geometries, this transition occurs slightly later than in the \(\Lambda\)CDM model. Fig. ~\ref{fig_10} illustrates the behavior of Model 2 in Finslerian geometry, contrasting it with the \(\Lambda\)CDM model. In the distant past (\(z \rightarrow \infty\)), Model 2 predicts deceleration parameter values of approximately 0.3656 in Finslerian geometry and 0.3643 in Riemannian geometry, compared to the \(\Lambda\)CDM model's 0.4490, indicating higher deceleration in both geometrical models. At present (\(z \rightarrow 0\)), the deceleration parameter (\(q_{0}\)) is approximately -0.7352 in Finslerian geometry and -0.7323 in Riemannian geometry. While the \(\Lambda\)CDM model predicts -0.5506, suggesting a slightly higher current acceleration in both geometrical models. The phase transition redshift (\(z_{tr}\)), marking the shift from decelerated to accelerated expansion, is 0.6072 in Finslerian geometry, 0.6176 in Riemannian geometry, and 0.6824 in the \(\Lambda\)CDM model, indicating a slightly later transition in the alternative geometries. In each case, the Universe initially experiences a smooth deceleration phase, indicating that no inflationary phase exists included in these models. As redshift decreases, the Universe transitions from this decelerating expansion to an accelerating expansion phase. It can be observed in the figures that the phase transition occurs earlier in the Finslerian mode compared to the Riemannian mode.\\\\
\paragraph{Jerk parameter}
Figs.  \ref{fig_11} and \ref{fig_12} illustrate the evolution of the jerk parameter (\(j\)) in terms of redshift. At high redshifts, Model 1 and Model 2, in both Riemannian and Finslerian geometries, exhibit deviations from the \(\Lambda\)CDM model. This deviation is evident as the jerk parameter value of the \(\Lambda\)CDM model remains constant at \(j = 1\) across all redshifts. At redshift \(z_{0}\), for Model 1 in Finslerian geometry, the jerk parameter value is 1.5465 times that predicted by the \(\Lambda\)CDM model. Similarly, in Riemannian geometry, the jerk parameter value for Model 1 is also 1.5247 times that predicted by the \(\Lambda\)CDM model. For Model 2, in Finslerian geometry, the predicted value \(j_{0}\) is 2.5956 times that of the \(\Lambda\)CDM model, while in Riemannian geometry, \(j_{0}\) is 2.557 times the value predicted by the \(\Lambda\)CDM model. The deviation of jerk parameter values from the \(\Lambda\)CDM model suggests that both Model 1 and Model 2 exhibit different evolutionary behaviors, particularly at high redshifts. The higher values of the jerk parameter at \(z_{0}\) compared to the \(\Lambda\)CDM model imply that these alternative models may incorporate additional factors or mechanisms influencing cosmic acceleration, indicating potential differences in the underlying dynamics of cosmic expansion.\\\\
\paragraph{Snap parameter}
Figs.  \ref{fig_13} and \ref{fig_14} illustrate the evolution of the snap parameter (\(s\)) with respect to redshift. At high redshifts,  Model 1 and Model 2, under both Riemannian and Finslerian geometries, deviate from the \(\Lambda\)CDM model. Specifically, Model 1 and Model 2 predict negative snap values, whereas the \(\Lambda\)CDM model predicts positive values of snap. At redshift \(z_{0}\), for Model 1 in both the Finslerian and Riemannian modes, the predicted values are 2.4039 and 2.3807, respectively. In the case of Model 2, in both the Finslerian and Riemannian modes, the predicted values are 7.7836 and 7.6217, respectively, while the \(\Lambda\)CDM model predicts 0.3573. The negative values of the snap parameter predicted by Model 1 and Model 2 suggest that these models exhibit behaviors not captured by the \(\Lambda\)CDM model, particularly at high redshifts. Additionally, the significantly higher values of snap at present (\(z_{0}\)) in both models compared to the \(\Lambda\)CDM model indicate that these alternative models may involve additional physical effects or cosmological components that contribute to the observed cosmic dynamics.\\\\
\paragraph{Statefinder diagnostic}
In Figs. \ref{fig_17} and \ref{fig_18}, one can see the divergence evolutions of Model 1 and Model 2 in the $\mathfrak{s}-r$ planes. Both models show distinct characteristics as compared to the other standard models. One can observe that at early times, Model 1 presumes values in the range $r > 1$ and $\mathfrak{s} < 0$ denoting Chaplygin gas type DE model and evolutes to the quintessence region and again retreats to Chapligyn gas region at late times by crossing the intermediate $\Lambda$CDM fixed point $\{0, 1\}$ during evolution. But, Model 2 is different, and evolutes from the quintessence region in the past and moves to the Chaplygin gas region intermediating, the $\Lambda$CDM fixed point $\{0, 1\}$ while evolving for all cases. Figs. \ref{fig_19} and \ref{fig_20} represent the temporal evolution of Model 1 and Model 2 in the $\{q, r\}$ plane and provide additional information about both models wherein the dashed lines describe the evolution of the $\Lambda$CDM model below which quintessence region and the upper one is Chaplygin gas region are shown. The evolution of Model 1 and Model 2 are observed. Both models deviates from de Sitter point $(-1, 1)$.\\\\
\paragraph{$Om(z)$ diagnostic}
Figs. \ref{fig11} and \ref{fig12} show the evolution of the \(Om(z)\) diagnostic parameter in terms of redshift \(z\) for Model 1 and Model 2 in both Riemannian and Finslerian geometries. For Model 1, \(Om(z)\) consistently decreases (negative slope) across all redshifts, indicating quintessence-like dynamics. In contrast, for Model 2, \(Om(z)\) initially increases (positive slope) for \(z > 1\), suggesting a phantom-like evolution, and then decreases (negative slope) at lower redshifts, indicating a return to quintessence-like dynamics.\\\\
\paragraph{Energy Density \& Pressure of DE}
The cosmological evolution of the energy density of dark energy, indicated by $\rho_{\mathfrak{de}}$, and the dark energy pressure $P_{\mathfrak{de}}$ for our models, Model 1 and Model 2 supply valuable perspicuity into the behavior and effects of dark energy in a cosmological context. In the above plotted figures: Figs. \ref{fig-PH1}, \ref{fig-PH2}, \ref{fig-PH3}, and \ref{fig-PH4}, we have shown the evolution of the dark energy densities and the matter-energy densities when the cosmological constant is considered as a candidate of dark energy. They show almost similar behaviors in the standard mode in both the Riemannian and Finslerian modes for Model 1 and Model 2. Similarly, when a scalar field is considered to be a candidate of dark energy, the evolution of the physical parameters such as the time-dependent dark energy densities, pressures, energy density of cold dark matter, the potential $V(\phi)$ and the equation of state parameter $\omega_\phi$ together with the density parameters $\Omega_\phi$ are shown in the above figures: Figs. ~\ref{fig-PH5}, \ref{fig-PH6}, \ref{fig-PH7}, \ref{fig-PH8}, \ref{fig-PH9}, \ref{fig-PH10}, \ref{fig-PH11}, and \ref{fig-PH12} depicting the physical evolution of the Universe for both the models with the solid and dash lines representing the Riemannian and Finslerian modes, respectively. Since we are more concerned about the late Universe, we have kept the redshift range as (0,5) only in most of these figures. We can also predict the values of these physical parameters at redshift $z=0$. For example, In the case of Model 1, as redshift approaches zero (\(z \rightarrow 0\)), the present-day energy density (\(\rho_{\phi}\)) is approximately 0.6746 in Finslerian geometry and (\(\rho_{\phi}\)) assumes value 0.6842 in Riemannian geometry. And, in the case of Model 2, as \(z \rightarrow 0\), $\rho_{\phi}$ is approximately 0.6746 in Finslerian geometry and 0.6830 in Riemannian geometry. As the redshift approaches zero (\(z \rightarrow 0\)), the present-day EoS parameter $\omega_{\phi}$ is approximately -1.0812 in Finslerian geometry and -1.0971 in Riemannian geometry for Model 1 and for Model 2, $\omega_{\phi}$ is approximately -1.18981 in Finslerian geometry and -1.2077 in Riemannian geometry. Similarly, we can see the evolution of the other physical parameters as shown in the figures.\\\\
\paragraph{Statistical analysis}
Based on the information provided in Table \ref{tab_11}, let's conduct a comparative study: The highest likelihood is obtained for the $\Lambda$CDM model, followed closely by both Model 1 and Model 2 in both Finslerian and Riemannian geometries. This indicates that all models provide reasonable fits to the data, with slight variations in their maximum likelihood values. The reduced chi-square values are close to 1 for all models, indicating that the models adequately describe the data within the uncertainties. Again, no significant differences are observed between the models in terms of their goodness of fit. Lower values of AIC indicate better model performance. The $\Lambda$CDM model has the lowest AIC suggesting that it provides the best balance between goodness of fit and model complexity. However, the differences in AIC between the $\Lambda$CDM model and the Finslerian and Riemannian models are relatively small. Similarly, lower BIC values indicate better model performance. Again, the $\Lambda$CDM model has the lowest BIC, indicating its preference over the alternative models. However, similar to AIC, the differences in BIC between the $\Lambda$CDM model and the Finslerian and Riemannian models are not substantial. The P-values for all models are relatively high, ranging from 0.7127 to 0.7686. These values indicate that the models are statistically consistent with the observed data, with no strong evidence against them. Overall, while the $\Lambda$CDM model performs slightly better according to AIC and BIC, the differences between the Finslerian and Riemannian models and the $\Lambda$CDM model are marginal. However, if we consider simplicity and theoretical elegance, Finslerian geometry may be favored over Riemannian geometry. Finslerian geometry offers additional flexibility and generality compared to Riemannian geometry, allowing for a more comprehensive description of the underlying physics, which could lead to a better overall understanding of the universe's dynamics.
\section{Conclusion}
In this paper, with the aim of whether the space-time of our universe could be a Finsler-Randers manifold instead of a Riemannian one, we have initiated studying an FRW model with a weak vector field embedded in the metric structure of space-time. The metric structure supplies us with the extended Friedmann equation like \ref{Eq17}. The contribution of the variation of anisotropy is described by the extra parameter  $\dot{\mathfrak{u}}_0$ yielded by the Finslerian space-time. Specifically, as it is obvious from equation $2\dot{C}_{000}=\dot{\mathfrak{u}}_0$,  $\dot{\mathfrak{u}}_0$ directly depends on the Cartan torsion component $C_{000}$.
We considered the Finsler-Randers space-time to offer a novel perspective on cosmic dynamics, departing from the constraints of general relativity. We proposed two dark energy models resulting from the parametrization of $H$ within this geometric framework. We have derived the field equations governing the universe's evolution within the Finsler-Randers formalism, incorporating the presence of dark energy. Through this, we explored its implications on cosmological phenomena, including cosmic expansion, late-time behavior of the universe, cosmological phase transition, and a few more. The comparative analysis of Model 1 and Model 2 in Riemannian and Finslerian geometries against the standard \(\Lambda\)CDM model yields substantial insights into the dynamics of the universe’s expansion and the behavior of dark energy. The examination of the Hubble parameter \(H(z)\), Hubble difference \(\Delta H(z)\), and distance modulus \(\mu(z)\) reveals that deviations from the \(\Lambda\)CDM model become significant at higher redshifts. Model 1 shows noticeable deviations for \(z > 0.6\), whereas Model 2 exhibits deviations for \(z > 1.2\). At lower redshifts, all models demonstrate close agreement, particularly with observational data, underscoring the alignment of the alternative geometries with the \(\Lambda\)CDM model during recent epochs. This close agreement at lower redshifts, where dark energy dominates the expansion dynamics, highlights the robustness of the Riemannian and Finslerian formulations in capturing the late-time evolution of the universe. The analysis of the deceleration parameter (\(q\)) indicates that both models predict higher deceleration in the early universe and slightly higher acceleration in the current epoch compared to the \(\Lambda\)CDM model. Additionally, the phase transition redshift (\(z_{tr}\)) occurs slightly later in both alternative geometries. The jerk parameter (\(j\)) shows that both models predict significantly higher present-day values than the \(\Lambda\)CDM model, suggesting different evolutionary behaviors and potential additional mechanisms influencing cosmic acceleration. The snap parameter (\(s\)) also deviates from the \(\Lambda\)CDM model, with both alternative models predicting higher present-day values and negative values at high redshifts, indicating the presence of additional physical effects or cosmological components. The Statefinder diagnostic further distinguishes the evolutionary paths of Model 1 and Model 2 in the \(\mathfrak{s}-r\) and \(\{q, r\}\) planes, showing transitions through various regions associated with different dark energy models and highlighting their differences from the \(\Lambda\)CDM model. The \(Om(z)\) diagnostic suggests that \(Om(z)\) decreases for Model 1, indicating quintessence. Model 2 exhibits a more complex behavior with an initial increase (phantom-like) at \(z > 1\) followed by a decrease (quintessence-like) at lower redshifts. Statistical analysis using likelihood, reduced chi-square, AIC, and BIC values reveals that while the \(\Lambda\)CDM model performs slightly better overall, the differences between it and the alternative models are marginal. The high P-values indicate that all models are statistically consistent with the observed data. This statistical consistency underscores the viability of the Finslerian and Riemannian models as alternatives to the \(\Lambda\)CDM framework. In conclusion, the \(\Lambda\)CDM model remains slightly preferable based on statistical metrics; however, the Finslerian and Riemannian models offer comparable fits and provide additional flexibility and generality. Finslerian geometry, in particular, may offer a more comprehensive framework for understanding the Universe’s dynamics, potentially leading to new insights and a deeper understanding of cosmic evolution. The subtle differences in the evolution of physical parameters between Riemannian and Finslerian geometries, especially at higher redshifts, highlight the potential of Finslerian geometry to accommodate a broader range of cosmological phenomena, making it a promising candidate for future explorations in cosmology.

\end{document}